%% file: arxiv.tex
\documentclass[11pt]{article}

\usepackage[a4paper,margin=1.1in]{geometry}

\usepackage{graphicx}
\usepackage{multirow}
\usepackage{amsmath,amssymb,amsfonts}
\usepackage{mathrsfs}
\usepackage[title]{appendix}
\usepackage{xcolor}
\usepackage{textcomp}
\usepackage{manyfoot}
\usepackage{booktabs}
\usepackage{algorithm}
\usepackage{algorithmicx}
\usepackage{algpseudocode}
\usepackage{listings}
\usepackage{subcaption}
\usepackage{siunitx}
\usepackage{tabularx}
\usepackage{comment}
\usepackage{wasysym}
\usepackage[section]{placeins}
\usepackage{orcidlink}
\usepackage{lineno}
\usepackage{microtype}
\usepackage{hyperref} % no options here
\hypersetup{
  colorlinks=true,
  linkcolor=blue,
  citecolor=blue,
  urlcolor=blue
}
\usepackage{cleveref}

\crefname{figure}{figure}{figures}
\Crefname{figure}{Figure}{Figures}

\crefname{table}{table}{tables}
\Crefname{table}{Table}{Tables}

\crefname{section}{section}{sections}
\Crefname{section}{Section}{Sections}

\crefname{appendix}{appendix}{appendices}
\Crefname{appendix}{Appendix}{Appendices}

\DeclareUnicodeCharacter{2212}{\ensuremath{-}}

\newcolumntype{C}{>{$}c<{$}} % centered math-mode column

\usepackage{array}
\newcolumntype{L}{>{$}l<{$}} % math-mode left-aligned column

\makeatletter
\newcommand*{\smallrel}[2][.8]{%
  \mathrel{\mathpalette{\smallrel@{#1}}{#2}}%
}
\newcommand*{\smallrel@}[3]{%
  \sbox0{$#2\vcenter{}$}%
  \dimen@=\ht0 %
  \raise\dimen@\hbox{%
    \scalebox{#1}{%
      \raise-\dimen@\hbox{$#2#3\m@th$}%
    }%
  }%
}
\makeatother

\title{Atmospheric neutrino constraints on Lorentz invariance violation with the first six detection units of KM3NeT/ORCA}

\author{%
{\large The KM3NeT collaboration}\\[0.8em]
{\small\texttt{
\href{mailto:km3net-pc@km3net.de}{km3net-pc@km3net.de}
\quad
\href{mailto:lukas.hennig@fau.de}{lukas.hennig@fau.de}
\quad
\href{mailto:alba.domi@fau.de}{alba.domi@fau.de}
}}
}
\begin{document}

\maketitle

\begin{abstract}
\input{sections/abstract}
\end{abstract}

% \flushbottom

% \tableofcontents

% \input{sections/proposed_structure}

\input{sections/introduction}

\input{sections/theory}

\input{sections/orca6detector}

\input{sections/data_sample_and_selection}

\input{sections/analysis_method}

\input{sections/results}

\input{sections/conclusion}

\clearpage
\section*{Acknowledgements}
\input{sections/acknowledgments}

\newpage

\bibliographystyle{JHEP}
\bibliography{references}

\input{sections/author_list}

\end{document}

%% file: sections/abstract.tex
Lorentz invariance is a fundamental symmetry underlying both the Standard Model of particle physics and General Relativity. Testing its validity provides a direct means of searching for new physics emerging near the Planck scale. A search for isotropic Lorentz invariance violation with 1.4 years of atmospheric neutrino data collected by a partial configuration of the KM3NeT/ORCA detector comprising six detection units is presented. No evidence for such violation is found; thus, competitive limits are set on a subset of isotropic Lorentz invariance violating coefficients, which complement and extend existing experimental constraints.

%% file: sections/introduction.tex
\section{Introduction}\label{introduction_sec}
Lorentz invariance is a cornerstone of modern fundamental physics and is
central to our understanding of the structure of spacetime. It underlies
both the Standard Model of particle physics and General Relativity, guaranteeing that the laws of physics take the same form for all inertial observers and that the outcomes of experiments are insensitive to the spatial orientation and uniform motion of the experimental setting. Despite its fundamental significance for the current understanding of spacetime, several new physics theories allow for violations of Lorentz invariance at or below the Planck scale, $M_\mathrm{P} \sim 10^{19}$ GeV. Examples of such theories include quantum gravity (QG) frameworks \cite{liv_qg} such as string theory \cite{liv_strings}, loop quantum gravity \cite{liv_lqg}, and doubly special relativity \cite{liv_dsr}, as well as noncommutative geometry \cite{noncommutative} and spacetime varying couplings \cite{st_varying_couplings}. Many of these models introduce a discretised structure of spacetime, which is inherently difficult to reconcile with continuous symmetries such as boost invariance. Testing the validity of Lorentz invariance across all experimentally accessible regimes could reveal new insights into the microscopic structure of spacetime and eventually impose stringent constraints on QG models.
\\
The Standard Model Extension (SME) is the most general effective field theory framework that consistently incorporates both the Standard Model and General Relativity, while allowing for general violations of both Lorentz symmetry and charge, parity, time (CPT) symmetry \cite{liv_kostelecky}. In the SME, Lorentz invariance violating effects are expressed through tensor operators of specific mass dimension, each contracted with corresponding coefficients that quantify the magnitude of the violation, which can be constrained experimentally.
In the neutrino sector, Lorentz invariance violation (LIV) can lead to modifications of standard oscillation behaviour, potentially producing observable deviations in the energy and zenith angle distributions of atmospheric and astrophysical neutrinos. Such effects provide a promising scenario for detection through high-precision neutrino telescope experiments. Previous searches have been carried out by Super-Kamiokande \cite{SK_paper} and IceCube \cite{IC2yrLIV,ic_astro}. No significant deviation from standard oscillations has been observed, and the results have therefore been used to place constraints on several SME coefficients.
\\
Atmospheric neutrino data collected over 1.4 years by a partial configuration of the KM3NeT/ORCA detector \cite{loi} comprising six detection units (indicated in the following as ORCA6) are used in this study to investigate potential LIV signatures. This is achieved through a detailed analysis of the energy-dependent spectra and angular distributions of the detected neutrino events.
\\
The paper is organised as follows: In \cref{sec:sme} the relevant theoretical background is presented, with an emphasis on aspects pertinent to ORCA6. In \cref{sec:detector} the KM3NeT/ORCA detector is described. The data sample collected with ORCA6 and used for this study is discussed in \cref{sec:data_sample_and_selection}. The analysis methodology employed in this work is detailed in \cref{sec:analysis_method}. Finally, the results are discussed in \cref{sec:results}.

%% file: sections/theory.tex
\section{Neutrino oscillations with isotropic Lorentz invariance violation}\label{sec:sme}
Within the SME, the general effective Hamiltonian for neutrino propagation with LIV includes four classes of operators, leading to a rich phenomenology that can be systematically tested \cite{liv_kostelecky}. The limit in which only LIV operators with mass dimension $d \leq 4$ are nonzero defines the renormalisable sector, which is known to be renormalisable at least at one loop and is expected to dominate at low energies \cite{kostelecky_renormalizable}. Non-renormalisable operators with $d \geq 5$ can also become relevant \cite{liv_kostelecky}: although the renormalisable SME sector suffices to describe physics at a low-energy scale, it becomes inadequate near high-energy scales where issues of causality and stability arise. In this regime, Planck‑suppressed higher‑dimension terms are expected to play an increasingly important role.
\\
This work specifically focuses on isotropic LIV which preserves rotational symmetry in a preferred reference frame. This class of models simplifies the testable parameter space, an approach that is particularly advantageous for neutrino oscillation experiments, as their sensitivity to energy and zenith angle distributions enables precise measurements of energy-dependent oscillation effects. Despite their simplicity, isotropic models retain sufficient complexity to provide meaningful opportunities for probing the fundamental structure of spacetime \cite{liv_kostelecky}.
\\
The evolution of a neutrino system in the presence of isotropic LIV is governed by the total Hamiltonian:
\begin{equation}\label{eq:fullH}
    H = U H_0 U^\dagger  + H_\mathrm{M} + H_{\mathrm{LIV}} \, ,
\end{equation}
where U denotes the PMNS mixing matrix \cite{Giuntibook}. The vacuum contribution is given by 
\begin{equation}\label{eq:standard_mixing}
    H_0 = \frac{1}{2E}
\begin{pmatrix}
0 & 0 & 0\\
0 & \Delta m_{21}^2 & 0\\
0 & 0 & \Delta m_{31}^2
\end{pmatrix}
\end{equation}
which describes standard oscillations in the neutrino mass basis. The matter interaction term 
\begin{equation}
    H_\mathrm{M} = \pm \sqrt{2}G_\mathrm{F} 
\begin{pmatrix}
N_e & 0 & 0\\
0 & 0 & 0\\
0 & 0 & 0
\end{pmatrix}
\end{equation}
is defined in the flavour basis and accounts for coherent forward scattering in ordinary matter. The LIV contribution is parametrised as
\begin{equation}\label{eq:hamiltonian}
    H_{\mathrm{LIV}} = \begin{pmatrix}
\mathring{a}_{ee}^{(3)} & \mathring{a}_{e\mu}^{(3)} & \mathring{a}_{e\tau}^{(3)}\\
\mathring{a}_{e\mu}^{(3)*} & \mathring{a}_{\mu\mu}^{(3)} & \mathring{a}_{\mu\tau}^{(3)}\\
\mathring{a}_{e\tau}^{(3)*} & \mathring{a}_{\mu\tau}^{(3)*} & \mathring{a}_{\tau\tau}^{(3)}
\end{pmatrix} - E \begin{pmatrix}
\mathring{c}_{ee}^{(4)} & \mathring{c}_{e\mu}^{(4)} & \mathring{c}_{e\tau}^{(4)}\\
\mathring{c}_{e\mu}^{(4)*} & \mathring{c}_{\mu\mu}^{(4)} & \mathring{c}_{\mu\tau}^{(4)}\\
\mathring{c}_{e\tau}^{(4)*} & \mathring{c}_{\mu\tau}^{(4)*} & \mathring{c}_{\tau\tau}^{(4)}
\end{pmatrix} + E^2 \mathring{a}_M^{(5)} - E^3 \mathring{c}_M^{(6)} + E^4 \mathring{a}_M^{(7)} - E^5 \mathring{c}_M^{(8)} + ...
\end{equation}
where \(\mathring{a}_{\alpha\beta}^{(d)}\) and \(\mathring{c}_{\alpha\beta}^{(d)}\), with \(\alpha,\beta\in\{e,\mu,\tau\}\), denote coefficients associated with CPT-odd and CPT-even isotropic LIV operators of mass dimension \(d\), respectively \cite{liv_kostelecky}. The symbols $\mathring{a}_M^{(d)}$, $\mathring{c}_M^{(d)}$ denote Hermitian matrices of isotropic LIV coefficients, analogous to the matrices for $d=3,4$ that are explicitly written out in \cref{eq:hamiltonian}. In the following, both cases are collectively denoted as \(\mathring{k}_{\alpha\beta}^{(d)}\); odd (even) values of \(d\) correspond to CPT-odd (CPT-even) operators. 
\\
In the SME Lagrangian density, which has mass dimension 4, each LIV term appears as the product of a coefficient and a LIV operator \cite{liv_kostelecky}. Therefore, if an operator has mass dimension \(d\), the corresponding coefficient \(\mathring{k}_{\alpha\beta}^{(d)}\) carries mass dimension \(4-d\). This is consistent with the \(E^{d-3}\mathring{k}_{\alpha\beta}^{(d)}\) terms in the resulting effective Hamiltonian in \cref{eq:hamiltonian}, where each contribution produces the correct mass dimension required by a Hamiltonian operator: \((d-3) + (4-d) = 1\).
\\
It is noteworthy that the matrix \( \mathring{a}_M^{(3)} \) closely resembles the structure of the propagation matrix in Non-Standard Interaction (NSI) models \cite{ORCA6_NSI}. Although a formal correspondence exists between NSI and LIV parameters, the two frameworks affect neutrino oscillations in fundamentally different ways \cite{tortola, Agarwalla:2023wft}. NSI effects require neutrino propagation through matter and depend on the matter potential, whereas LIV influences oscillation probabilities even in vacuum and remains independent of matter density variations.
\\
The LIV coefficients of a given mass dimension \(d\) can be grouped into the three off-diagonal coefficients \(\mathring{k}_{e\mu}^{(d)}, \mathring{k}_{e\tau}^{(d)}, \mathring{k}_{\mu\tau}^{(d)}\) and the three diagonal coefficients \(\mathring{k}_{ee}^{(d)}, \mathring{k}_{\mu\mu}^{(d)}, \mathring{k}_{\tau\tau}^{(d)}\). As described in \cref{sec:data_sample_and_selection}, the selected ORCA6 event sample is dominated by $\nu_\mu$-induced events and probes atmospheric neutrinos propagating across baselines up to the Earth's diameter $D_\oplus$. The following two subsections therefore focus on the impact of isotropic LIV on the muon-neutrino disappearance channel, \(P(\nu_\mu\rightarrow\nu_\mu)\), in the energy range relevant for this analysis.

\subsection{Off-diagonal LIV coefficients}
The complex-valued LIV coefficients \(\mathring{k}_{e\mu}^{(d)}, \mathring{k}_{e\tau}^{(d)}, \mathring{k}_{\mu\tau}^{(d)}\) appear on the off-diagonal of $H_\mathrm{LIV}$ in \cref{eq:hamiltonian}. Each of these coefficients $\mathring{k}_{\alpha\beta}^{(d)}$ induces an additional mixing among the flavour states $\alpha \leftrightarrow \beta$, which is added to the mixing introduced by the off-diagonal terms in the standard-oscillation Hamiltonian $U H_0 U^\dagger$ in the flavour basis. For simplicity, in the following discussion it is assumed that only one off-diagonal coefficient $\mathring{k}_{\alpha\beta}^{(d)}$ is nonzero at a time.
\\
At sufficiently high energies, where the LIV-induced mixing \(E^{d-3}\mathring{k}_{\alpha\beta}^{(d)}\) dominates the standard mixing driven by $\Delta m_{ij}^2/2E$ (see \cref{eq:standard_mixing}), a nonzero coefficient $\mathring{k}_{\alpha\beta}^{(d)}$ introduces transitions between the flavours $\alpha \leftrightarrow \beta$ that are absent in the standard oscillation scenario. In this high-energy limit, the oscillation phase scales as \cite{IC2yrLIV}
\begin{equation}
\phi_{\mathrm{LIV}}(E) \; \propto\; L\,E^{d-3}\,|\mathring{k}_{\alpha\beta}^{(d)}|.
\label{eq:phi_liv_scaling}
\end{equation}
The first minimum in the neutrino survival probability \(P(\nu_\alpha\rightarrow\nu_\alpha)\) occurs
when \(\phi_{\mathrm{LIV}}= \pi/2\). For \(d>3\), this implies that
the characteristic energy scale of the LIV-induced oscillations shifts to
higher energies as \(|\mathring{k}_{\alpha\beta}^{(d)}|\) decreases.
\begin{figure}
\centering
  \includegraphics[width=1.0\linewidth]{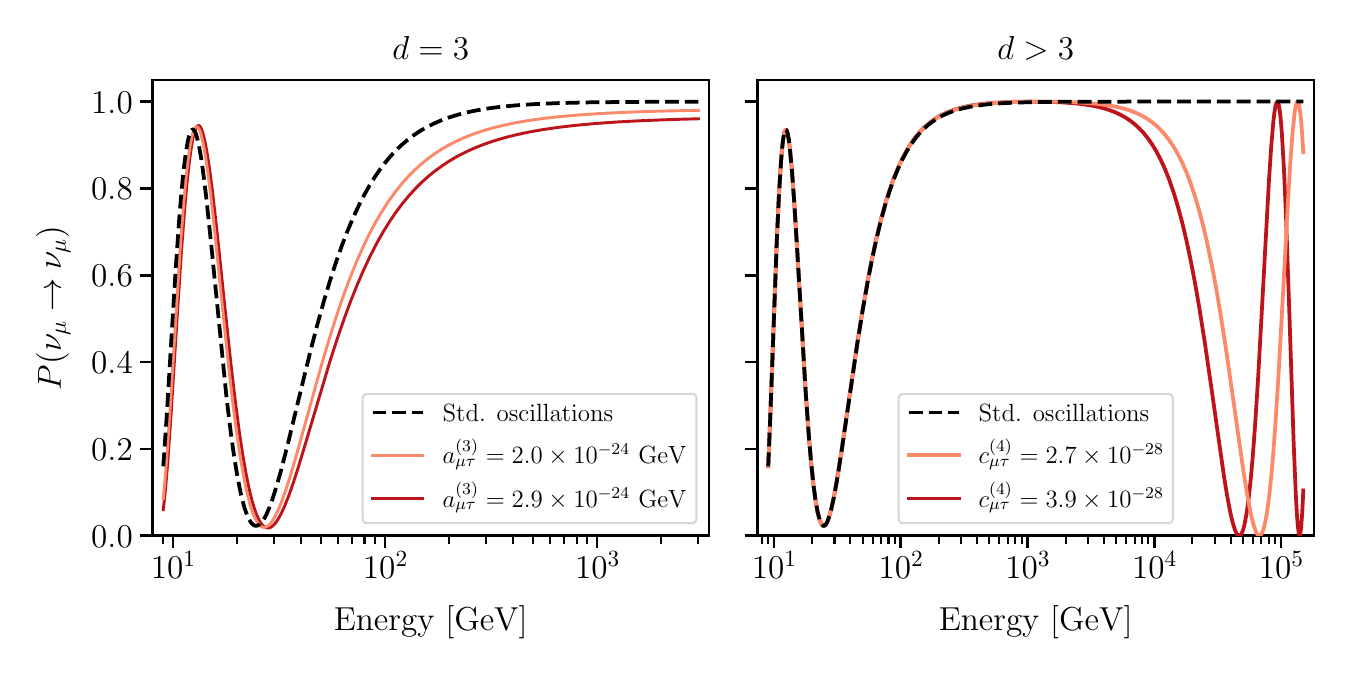}
  \caption{Muon neutrino survival probability as a function of neutrino energy. The neutrino is assumed to have traversed the Earth with a baseline $L \simeq D_{\oplus}$.  The effect of mass dimension $d = 3$ coefficients is illustrated in the left panel using different values of $\mathring{a}_{\mu\tau}^{(3)}$ as an example. The $d>3$ case is shown in the right panel, with $\mathring{c}_{\mu\tau}^{(4)}$ being chosen as a representative example. The values for $\mathring{a}_{\mu\tau}^{(3)}$ and $\mathring{c}_{\mu\tau}^{(4)}$ are the 90\% (orange) and 99\% (red) confidence level (CL) constraints reported by the IceCube collaboration in \cite{IC2yrLIV}. The oscillation probabilities are computed using OscProb \cite{oscprob}.}
  \label{fig:Pmumu_offdiag_different_mass_dim}
\end{figure}
This behaviour is illustrated in \cref{fig:Pmumu_offdiag_different_mass_dim}
for a neutrino traversing the Earth (\(L\simeq D_\oplus\)), using the
\(\mu\tau\) sector as an example. For \(d=3\) (left panel), the survival probability \(P(\nu_\mu\rightarrow\nu_\mu)\) approaches a constant value at high energies because the phase $\phi_{\mathrm{LIV}}$ in \cref{eq:phi_liv_scaling} becomes energy independent. Such a predicted, nearly constant, $\nu_\mu$ deficit is partly degenerate with
uncertainties on the flux normalisations of atmospheric and cosmic neutrinos, which can reduce the sensitivity of high-energy neutrino telescopes like IceCube or KM3NeT/ARCA to the mass dimension 3 off-diagonal coefficients. In contrast, the KM3NeT/ORCA detector is optimised for the detection of neutrinos with energies as low as a few $\mathrm{GeV}$ \cite{loi}, where standard oscillations induce an energy-dependent structure. In this region, a nonzero \(\mathring{a}_{\alpha\beta}^{(3)}\) produces characteristic shape
distortions that are exploited in this analysis, leading to stringent limits on these coefficients.
\\
For \(d>3\) (right panel), the phase $\phi_{\mathrm{LIV}}$ grows with energy and the LIV-induced oscillations move to higher energies, making the study of these coefficients better suited for analyses with higher-energy neutrino samples
using, e.g, IceCube \cite{IC2yrLIV,ic_astro} or KM3NeT/ARCA. For these reasons, the analysis presented in this paper focuses on constraining the three off-diagonal dimension-3 coefficients \(|\mathring{a}_{e\mu}^{(3)}|\), \(|\mathring{a}_{e\tau}^{(3)}|\), and \(|\mathring{a}_{\mu\tau}^{(3)}|\). 

\subsection{Diagonal LIV coefficients}
The real-valued diagonal part of the Hamiltonian $H_{\mathrm{LIV}}$ can be written as
\begin{equation}
    H_{\mathrm{LIV}} = E^{d-3}\begin{pmatrix}
        \mathring{k}_{ee}^{(d)} & 0 & 0 \\
        0 & \mathring{k}_{\mu\mu}^{(d)} & 0 \\
        0 & 0 & \mathring{k}_{\tau\tau}^{(d)}
    \end{pmatrix}.
    \label{eq:hliv}
\end{equation}
Oscillation probabilities are invariant under any transformation $H \rightarrow H-\lambda(E) \, \mathbb{I}$, where $\mathbb{I}$ is the identity matrix, since this transformation only adds a global phase. Hence, it is allowed to subtract the term $E^{d-3}\mathring{k}_{\tau\tau}^{(d)}$ from the diagonal of the Hamiltonian. This leaves two independent diagonal combinations,
\begin{equation}
    H_{\mathrm{LIV}} = E^{d-3}
    \begin{pmatrix}
        \mathring{k}_{ee}^{(d)} - \mathring{k}_{\tau\tau}^{(d)} & 0 & 0 \\
        0 & \mathring{k}_{\mu\mu}^{(d)} - \mathring{k}_{\tau\tau}^{(d)} & 0 \\
        0 & 0 & 0
    \end{pmatrix}.
    \label{eq:diagmatrix}
\end{equation}
After including matter effects relevant for atmospheric neutrinos traversing the Earth, the interaction Hamiltonian in the flavour basis becomes
\begin{equation}
    H_\mathrm{I} \;=\; H_\mathrm{M} + H_{\mathrm{LIV}}
    \;=\;
    \begin{pmatrix}
        V_e + E^{d-3}\,(\mathring{k}_{ee}^{(d)}-\mathring{k}_{\tau\tau}^{(d)}) & 0 & 0 \\
        0 & E^{d-3}\,(\mathring{k}_{\mu\mu}^{(d)}-\mathring{k}_{\tau\tau}^{(d)}) & 0 \\
        0 & 0 & 0
    \end{pmatrix}.
    \label{eq:Hi}
\end{equation}
\begin{figure}[]
    \centering
    \includegraphics[width=\linewidth]{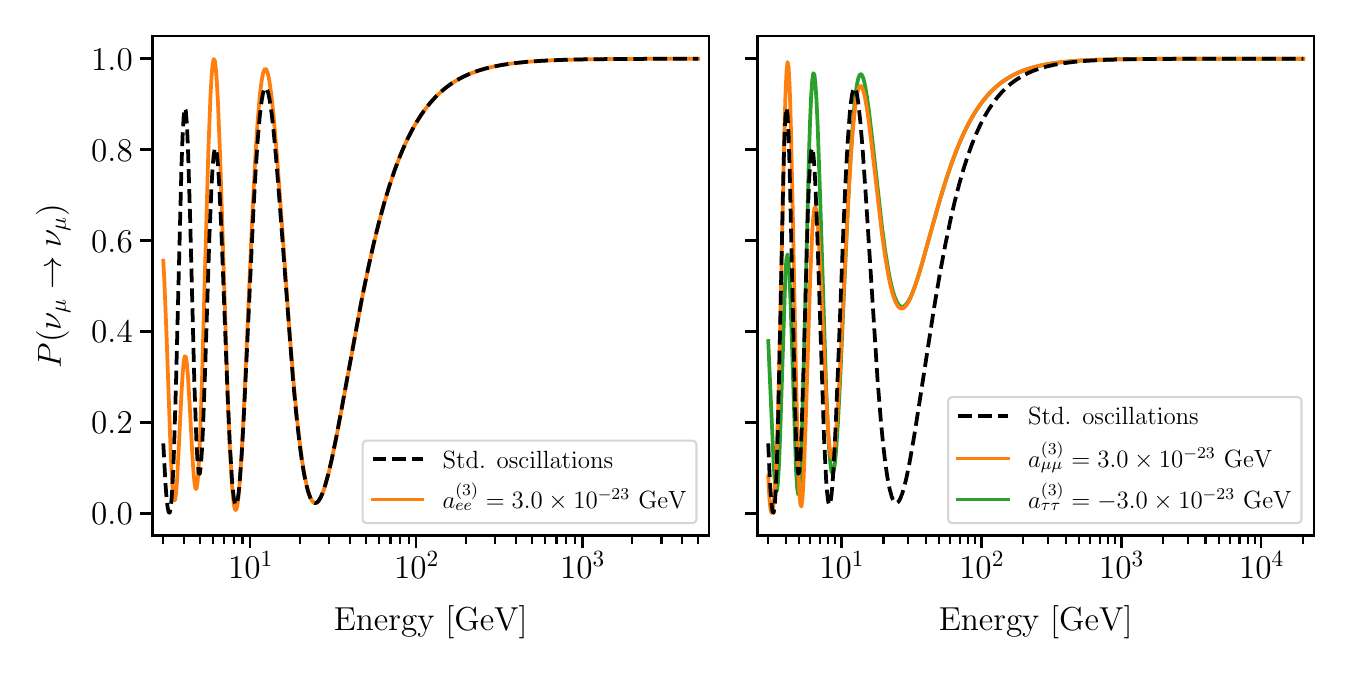}
    \caption{Muon neutrino survival probability as a function of neutrino energy. The neutrino is assumed to have traversed the Earth with a baseline $L \simeq D_{\oplus}$. The left panel shows the effect of a nonzero $\mathring{k}_{ee}^{(d)}$, while the right panel illustrates the effect of a nonzero $\mathring{k}_{\mu\mu}^{(d)}$ and $- \mathring{k}_{\tau\tau}^{(d)}$, respectively. Mass dimension $d=3$ has been chosen as a representative example, using a coefficient value suitable for an illustration of LIV effects. The oscillation probabilities are computed using OscProb \cite{oscprob}.}
    \label{fig:Pmumu_diagonal_mumu_tautau_ee}
\end{figure}
\\
First, a nonzero $\mathring{k}_{ee}^{(d)}$ with $\mathring{k}_{\mu\mu}^{(d)}=\mathring{k}_{\tau\tau}^{(d)}=0$ is considered. Equation~\ref{eq:Hi} shows that $\mathring{k}_{ee}^{(d)}$ enters as an additive contribution to the $ee$ entry of the Hamiltonian matrix, i.e., it effectively shifts the standard matter potential $V_e$. Since the contribution scales as $E^{d-3}$, the effective shift of $V_e$ is energy independent for $d=3$ and energy dependent for $d>3$. As illustrated in \cref{fig:Pmumu_diagonal_mumu_tautau_ee} (left panel), a nonzero $\mathring{k}_{ee}^{(d)}$ mainly affects $P(\nu_\mu\rightarrow\nu_\mu)$ in the few-GeV region, where it produces both positive and negative deviations with respect to the standard oscillation curve. After folding in the finite energy resolution, these deviations largely average out, and the limited statistics corresponding to the ORCA6 exposure does not allow for resolving the remaining effect. Therefore, $\mathring{k}_{ee}^{(d)}$ is not constrained in this analysis.
\\
Next, a nonzero $\mathring{k}_{\mu\mu}^{(d)}$ with $\mathring{k}_{ee}^{(d)}=\mathring{k}_{\tau\tau}^{(d)}=0$ is considered. In this case, $\mathring{k}_{\mu\mu}^{(d)}$ acts as an additional potential in the $\mu\mu$ entry of \cref{eq:Hi}. As shown in \cref{fig:Pmumu_diagonal_mumu_tautau_ee} (right panel), a nonzero $\mathring{k}_{\mu\mu}^{(d)}$ produces a significant modification of $P(\nu_\mu\rightarrow\nu_\mu)$ in the energy range relevant for this analysis. Therefore, the analysis is sensitive to the LIV effects induced by $\mathring{k}_{\mu\mu}^{(d)}$.
\\
Finally, a nonzero $\mathring{k}_{\tau\tau}^{(d)}$ with $\mathring{k}_{ee}^{(d)}=\mathring{k}_{\mu\mu}^{(d)}=0$ is considered. As can be seen in \cref{eq:Hi}, a nonzero $\mathring{k}_{\tau\tau}^{(d)}$ simultaneously induces an effective negative shift in both the $ee$ and $\mu\mu$ entries, i.e. it combines the qualitative features discussed above for $\mathring{k}_{ee}^{(d)}$ and $\mathring{k}_{\mu\mu}^{(d)}$. In the energy range relevant for this analysis, the dominant impact on $P(\nu_\mu\rightarrow\nu_\mu)$ arises from the effective shift of the $\mu\mu$ entry, while the accompanying shift of the $ee$ entry mainly affects the few-GeV region. Consequently, the effects of varying $\mathring{k}_{\mu\mu}^{(d)}$ and of varying $-\mathring{k}_{\tau\tau}^{(d)}$ appear nearly degenerate in \cref{fig:Pmumu_diagonal_mumu_tautau_ee} (right panel) at energies $\mathord{>}\,10\,\mathrm{GeV}$. For this reason, constraints on the diagonal combinations $(\mathring{k}_{\mu\mu}^{(d)}-\mathring{k}_{\tau\tau}^{(d)})$ are reported, which for brevity are denoted as $\mathring{k}_{\mu\mu-\tau\tau}^{(d)}$.
\\
These diagonal combinations enter the Hamiltonian as an effective potential term proportional to $E^{d-3}\,\mathring{k}_{\mu\mu-\tau\tau}^{(d)}$. Since the energy dependence becomes stronger with increasing $d$, the same effective potential size in a fixed energy range can be produced by progressively smaller coefficient values as $d$ increases. This scaling behaviour explains why the ORCA6 data sample retains sensitivity to diagonal coefficients of higher mass dimension. Consequently, the diagonal combination $\mathring{k}_{\mu\mu-\tau\tau}^{(d)}$ is constrained from mass dimension $d = 3$ up to $d=8$.

%% file: sections/orca6detector.tex
\section{The KM3NeT/ORCA detector}\label{sec:detector}

The KM3NeT collaboration is constructing the KM3NeT/ORCA detector (Oscillation Research with Cosmics in the Abyss) \cite{loi}, a deep-sea water-Cherenkov neutrino telescope positioned 40\,km offshore Toulon, France, at a depth of 2450\,m below sea level. KM3NeT/ORCA is specifically designed for the measurement of atmospheric neutrino oscillations in the $(1{-}100)\,\mathrm{GeV}$ energy range.
\\
When neutrinos interact within or near the instrumented volume, they generate secondary charged particles that induce Cherenkov radiation, which is detected by a three-dimensional array of digital optical modules \cite{DOM_paper}. Each module houses 31 photomultiplier tubes (PMTs) alongside readout electronics and calibration sensors for precise positioning, orientation, and timing. The optical modules are arranged vertically along detection units each comprising 18\,modules, which are anchored to the seabed and maintained in approximate vertical position by buoyancy distributed in the optical modules and a top buoy, enabling accurate reconstruction of neutrino energy and arrival direction. KM3NeT/ORCA is deployed in a modular configuration with detection units spaced horizontally at approximately 20\,m intervals and optical modules separated vertically by 9\,m along each unit. The completed detector will feature more than a hundred detection units, creating a dense optical array optimised for high-precision atmospheric neutrino oscillation measurements and neutrino mass ordering determination.

%% file: sections/data_sample_and_selection.tex
\section{Data sample and selection}\label{sec:data_sample_and_selection}
The data used in this analysis were collected with ORCA6, which refers to the configuration of the KM3NeT/ORCA detector from January 2020 to November 2021. The collected data were filtered according to quality criteria for environmental conditions and the stability of the data-taking, resulting in a dataset with 510 days of livetime and 433 kton-years of exposure \cite{Victor_standard_osc_paper}.
\\
The deep-sea environment provides a constant background of optical photons from the radioactive decay of $\,{}^{40}\mathrm{K}$ and from bioluminescence caused by microorganisms. This dominating background is reduced by applying trigger algorithms to search for causal correlations between the measured PMT signals. The remaining background events due to random coincidences are further suppressed to a negligible sub-percent level by applying selection cuts on quality criteria based on the output of reconstruction algorithms.
\\
At this stage of the selection, muons created in the atmosphere dominate the sample by seven orders of magnitude over the neutrino events. The muon contamination is reduced in two steps. The first step exploits the fact that the Earth is transparent for neutrinos at the energies considered in this analysis, while it is opaque to muons. By only selecting events reconstructed as up-going in the detector, the Earth is used as a shield against the atmospheric muon flux. Furthermore, a boosted decision tree (BDT) is trained to remove remaining muon events that are misreconstructed as up-going, allowing for a reduction of the muon contamination below 2\% while retaining 60\% of the estimated neutrino signal of the previous step.
\\
The last selection level differentiates between the two main event topologies, which are referred to as track-like and shower-like events. Track-like events are predominantly caused by muon neutrinos undergoing a charged-current (CC) interaction resulting in the creation of a muon. This charged particle induces the emission of Cherenkov radiation in the water while propagating tens to hundreds of metres, resulting in an elongated, track-like light signal in the detector. In contrast, shower-like events are mainly caused by CC interactions of electron neutrinos and tau neutrinos as well as neutral-current (NC) interactions induced by all flavours. These interactions produce electromagnetic and hadronic showers emitting light along a distance which is small compared to the distance between optical modules, resulting in the measurement of a nearly spherically-distributed light signal. The discrimination between track- and shower-like events is performed with a second BDT. Based on the scores of both BDTs, three event classes are defined, which are termed High Purity Tracks (HPT), Low Purity Tracks (LPT), and Showers. These classes are optimised for sensitivity on standard oscillation parameters and are used for all ORCA6 analyses \cite{Victor_standard_osc_paper,ORCA6_sterile,ORCA6_decoherence,ORCA6_neutrino_decay,ORCA6_NSI,ORCA6_tau}. The best fit to the standard oscillation hypothesis predicts 5828 observed events summed over the three classes. The muon neutrino and antineutrino fraction is 95\% for the HPT class, 90\% for the LPT class, and 46\% for the Showers class, while the atmospheric muon contaminations are 0.4\%, 4\%, and 1\%, respectively. Further details on the data sample used in this analysis, including its energy distribution and reconstruction resolutions of ORCA6, can be found in \cite{Victor_standard_osc_paper}.

%% file: sections/analysis_method.tex
\section{Analysis method}\label{sec:analysis_method}
\subsection{Likelihood definition and confidence interval construction}\label{sec:likelihood_def}
The detected events in each class are binned in the logarithm of the reconstructed energy, $\log_{10}(E_{\mathrm{reco}}/\mathrm{GeV})$, and  in the reconstructed cosine of the zenith angle, $\cos\theta$. Ten equally-spaced bins are used for the reconstructed $\cos\theta \in [\,-1,0\,]$, with $-1$ ($0$) representing vertically up-going (horizontal) events. The reconstructed energy is binned into 15 log-spaced bins with $E_\mathrm{reco} \in [\,2\,\mathrm{GeV},\, 1\,\mathrm{TeV}\,]$.
\\
The histograms obtained from observed events are compared to templates obtained using Monte Carlo (MC) simulated events that are reweighted according to the hypothesis under test. 
Point estimates for the LIV-parameters are obtained through Maximum Likelihood Estimation. The negative log-likelihood used in this analysis is
\begin{equation}
-2 \log \mathcal{L} = 
\sum_{i} 2 \left[
\left( \beta_i N^{\mathrm{exp}}_{i} - N^{\mathrm{obs}}_{i} \right)
+ N^{\mathrm{obs}}_{i} \log \left( \frac{N^{\mathrm{obs}}_{i}}{\beta_i N^{\mathrm{exp}}_{i}} \right)
\right]
+ \frac{(\beta_i - 1)^2}{\sigma_{\beta_i}^2}
+ \sum_{k} \left( \frac{\eta_k - \langle \eta_k \rangle}{\sigma_{\eta_k}} \right)^2 .
\end{equation}
The index $i$ in the first sum runs over all the aforementioned bins in reconstruction space, computing the Poisson likelihood that a bin with $N^{\mathrm{exp}}$ expected events has $N^{\mathrm{obs}}$ observed events. The second term treats the limited MC statistics by introducing normally-distributed coefficients $\beta_i$, following the Barlow and Beeston light method \cite{BARLOW1993219,Conway:1333496}. The index $k$ in the last sum runs over nuisance parameters, which treat systematic uncertainties from various sources. They are further discussed in \cref{subsection:nuisance_params}. The available information on the range of a nuisance parameter $\eta_k$ is modelled using a Gaussian prior with mean $\langle \eta_k \rangle$ and standard deviation $\sigma_{\eta_k}$. In the case of no available information, the nuisance parameter is allowed to vary freely and does not enter the sum over the index $k$.
\\
In order to construct confidence intervals, a test statistic (TS) has to be chosen. This analysis uses the likelihood ratio defined as
\begin{equation}
-2 \Delta \log \mathcal{L}(\omega_0) 
= -2 \log \left( 
\frac{\mathcal{L}(\omega_0, \hat{\hat{\vec{\eta}}})}
{\mathcal{L}(\hat{\omega}, \hat{\vec{\eta}})}
\right), \label{eq:likelihood_ratio}
\end{equation}
where $\omega$ denotes the LIV parameter that is allowed to vary and $\vec{\eta}$ denotes the vector of nuisance parameters. The hat over $\hat{\omega}$ and $ \hat{\vec{\eta}}$ denotes the point estimate that minimises the negative log-likelihood. The implementation of this minimisation procedure is based on the MINUIT library \cite{MINUIT}. For the construction of confidence intervals, a grid of LIV parameter values is scanned. For each LIV parameter value $\omega_0$ on this grid, the negative log-likelihood $-2\log \mathcal{L}(\omega_0, \vec{\eta})$ is minimised with $\omega_0$ kept fixed, resulting in a profiled best fit vector $\hat{\hat{\vec{\eta}}}$ for the nuisance parameters. According to Wilks' theorem \cite{Wilks}, the distribution of this TS can be approximated by a $\chi^2$-distribution with one degree of freedom, which is used for the construction of confidence intervals. A Feldman-Cousins construction \cite{Feldman-Cousins} is not used here, since the required pseudo-experiment ensembles for each scanned LIV coefficient would be computationally too costly given the large number of parameters studied in this analysis.

\subsection{Nuisance parameters}\label{subsection:nuisance_params}
Fifteen nuisance parameters are used to handle systematic uncertainties arising from oscillation parameters, cross sections and selection efficiencies, detector characteristics, and the atmospheric neutrino flux. A complete list of the nuisance parameters can be found in \cref{tab:nuisance_params}. This section gives a brief overview of the nuisance parameters, with a comprehensive explanation available in \cite{Victor_standard_osc_paper}.
\\
The uncertainties on standard oscillation parameters are taken into account by letting $\Delta m^2_{31}$ and $\theta_{23}$ vary freely during the fit. Other oscillation parameters have a negligible effect on the analysis and thus are fixed to their best fit values from NuFit 5.0 with Super-Kamiokande data assuming normal ordering \cite{nufit5.0}.
\\
The overall normalisation factor $f_{\mathrm{all}}$ is applied as a common scaling factor to the number of expected selected events in all classes. In addition, the relative normalisation factors $f_{\mathrm{HPT}}$ and $f_{\mathrm{S}}$ are applied to scale the numbers of expected selected events in the HPT and Shower classes, respectively, relative to this global scaling. This parameterisation allows for a combined treatment of the uncertainties in the neutrino cross sections and the selection efficiencies. The uncertainty in the estimation of the remaining atmospheric muon background is modelled using a normalisation $f_\mu$. Event selection uncertainties of CC $\nu_\tau$ events and all-flavour NC events are analogously modelled using scaling factors $f_{\tau\mathrm{CC}}$ and $f_\mathrm{NC}$ applied to these respective types of events.
\\
A nuisance parameter $E_s$ is introduced to shift the absolute energy scale of the detector, which accounts for uncertainties in the optical properties of the detector medium, the PMT efficiencies, and hadronic showers. Furthermore, the scaling factor $f_\mathrm{HE}$ is applied to high-energy events, accounting for different assumptions made during the simulated light propagation of low-energy and high-energy events in seawater.
\\
The nominal atmospheric neutrino flux used in this analysis is the HKKMS 2015 model \cite{HKKMS2015} averaged over the azimuth and evaluated at the Fr\'ejus site. The spectral index $\delta_\gamma$ models uncertainties in the steepness of the energy spectrum, while the parameter $\delta_\theta$ relates to uncertainties in the zenith distribution of the atmospheric neutrino flux. The remaining three nuisance parameters model uncertainties in the $\nu/\bar{\nu}$ and $\nu_\mu / \nu_e$ ratios.

\begin{table}
\centering
% --- spacing tweaks ---
\setlength{\tabcolsep}{4pt}       % horizontal cell padding (default ~6pt)
\renewcommand{\arraystretch}{1.2} % vertical row spacing (default 1.0)

\begin{tabular}{@{} p{0.5\linewidth} r c @{}}
\toprule
\multicolumn{2}{l}{\textbf{Parameter}} & \textbf{Uncertainty} \\
\midrule
\multicolumn{3}{@{}l}{\textit{Oscillation parameters:}} \\
\hspace{1em}Mass-squared difference      & $\Delta m^2_{31}$    & ---      \\
\hspace{1em}Mixing angle                 & $\theta_{23}$        & ---      \\
\midrule
\multicolumn{3}{@{}l}{\textit{Normalisations:}} \\
\hspace{1em}High-purity tracks           & $f_{\mathrm{HPT}}$   & ---      \\
\hspace{1em}Showers                      & $f_{\mathrm{S}}$     & ---      \\
\hspace{1em}Overall                      & $f_{\mathrm{all}}$   & ---      \\
\hspace{1em}Atm.\ muon background        & $f_{\mu}$            & ---      \\
\hspace{1em}$\nu_{\tau}+\bar{\nu}_{\tau}$ CC normalisation & $f_{\tau\mathrm{CC}}$ & $\pm 20\%$ \\
\hspace{1em}NC normalisation             & $f_{\mathrm{NC}}$    & $\pm 20\%$ \\
\midrule
\multicolumn{3}{@{}l}{\textit{Detector characteristics:}} \\
\hspace{1em}Energy scale                 & $E_s$                & $\pm 9\%$  \\
\hspace{1em}High-energy light yield      & $f_{\mathrm{HE}}$    & $\pm 50\%$ \\
\midrule
\multicolumn{3}{@{}l}{\textit{Atmospheric neutrino flux:}} \\
\hspace{1em}Spectral index               & $\delta_{\gamma}$    & $\pm 0.3$ \\
\hspace{1em}Ratio up-going to horizontal $\nu$ & $\delta_{\theta}$ & $\pm 2\%$ \\
\hspace{1em}$\nu_{\mu}/\bar{\nu}_{\mu}$ ratio  & $s_{\mu\bar{\mu}}$ & $\pm 5\%$ \\
\hspace{1em}$\nu_{e}/\bar{\nu}_{e}$ ratio      & $s_{e\bar{e}}$     & $\pm 7\%$ \\
\hspace{1em}$\nu_{\mu}/\nu_{e}$ ratio          & $s_{e\mu}$         & $\pm 2\%$ \\
\bottomrule
\end{tabular}

\vspace{0.25cm}
\caption{The nuisance parameters and their prior uncertainties as defined in \cite{Victor_standard_osc_paper}. Parameters without a value in the ``Uncertainty'' column are allowed to vary freely during the fit.}
\label{tab:nuisance_params}
\end{table}

%% file: sections/results.tex
\section{Results} \label{sec:results}
The analysis has been performed individually for all off-diagonal coefficients of mass dimension 3 and for the diagonal coefficient combinations $\mathring{k}^{(d)}_{\mu\mu} - \mathring{k}^{(d)}_{\tau\tau}$ with $ 3 \leq d \leq 8$, i.e., only one LIV coefficient is allowed to be nonzero in each fit. No significant LIV effect has been found. Hence, in this section the upper limits obtained with the ORCA6 data sample are reported and compared to limits set by other experiments.

\subsection{Off-diagonal coefficients}
\begin{figure}[!htbp]
\centering

% --- top row ---
\begin{subfigure}[t]{0.49\linewidth}
  \centering
  \includegraphics[width=\linewidth]{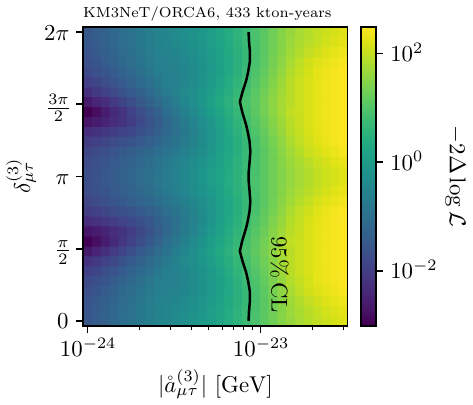}
\end{subfigure}\hfill
\begin{subfigure}[t]{0.49\linewidth}
  \centering
  \includegraphics[width=\linewidth]{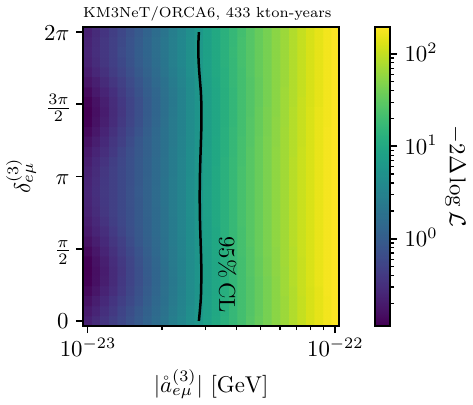}
\end{subfigure}

\vspace{0.6em}

% --- bottom row (centered) ---
\begin{subfigure}[t]{0.49\linewidth}
  \centering
  \includegraphics[width=\linewidth]{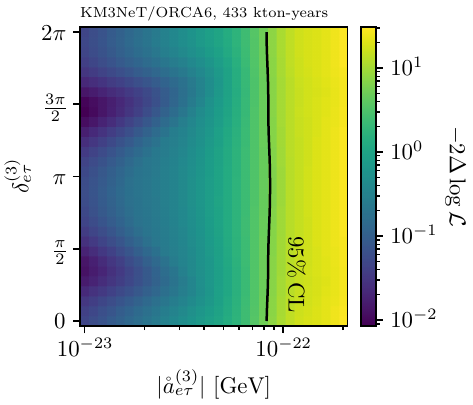}
\end{subfigure}

\caption{Two-dimensional sensitivities for the magnitude and phase of $d=3$ off-diagonal coefficients. The 95\% confidence level contour is shown in each panel.}
\label{fig:2d_sensitivities}
\end{figure}

In the SME, the off-diagonal coefficients $\mathring{a}_{\alpha\beta}^{(3)}$ are complex and can be written as
$\mathring{a}_{\alpha\beta}^{(3)} = |\mathring{a}_{\alpha\beta}^{(3)}| \, e^{i\delta_{\alpha\beta}^{(3)}}$.
The impact of the complex phase has been assessed by performing two-dimensional sensitivity scans using the test statistic in \cref{eq:likelihood_ratio}.
Assuming Wilks' theorem \cite{Wilks}, the test statistic follows a $\chi^2$ distribution with two degrees of freedom, which is used to draw the 95\% CL contour in \cref{fig:2d_sensitivities}. Along the 95\% CL contour, the inferred upper limit on $|\mathring{a}_{\mu\tau}^{(3)}|$ varies at the $\sim 15\%$ level when scanning over the phase $\delta_{\mu\tau}^{(3)}$ (upper left panel). This dependence is sufficiently weak that the analysis could be simplified to one-dimensional scans by fixing the phase. For $|\mathring{a}_{e\mu}^{(3)}|$ and $|\mathring{a}_{e\tau}^{(3)}|$ (upper right and bottom panel), the corresponding variation along the 95\% CL contour is only at the few-percent level, so the choice of a fixed phase has a negligible impact within the sensitivity of this dataset. Therefore, $\delta_{\alpha\beta}^{(3)}=0$ is fixed for all off-diagonal coefficients and only the magnitudes $|\mathring{a}_{\alpha\beta}^{(3)}|$ are constrained. For $\mu\tau$ this corresponds to a conservative choice, while for $e\mu$ and $e\tau$ the effect of the phase choice is negligible.
\begin{figure}[!htbp]
\centering
  \includegraphics[width=0.75\linewidth]{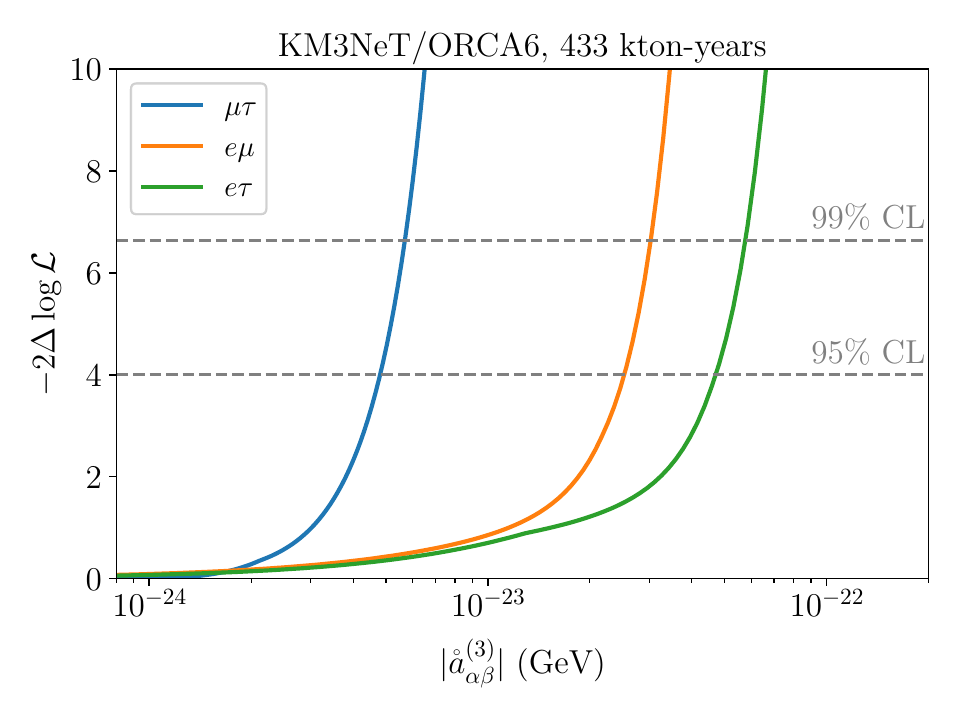}
  \caption{Likelihood ratio scans for $d=3$ off-diagonal coefficients. The intersections with the dashed lines define the upper limits.}
  \label{fig:off_diagonal_mass_dim_3_lr_scans}
\end{figure}
\\
Likelihood ratio scans of the three off-diagonal coefficients are shown in \cref{fig:off_diagonal_mass_dim_3_lr_scans}. The derived 95\% and 99\% CL upper limits and best-fit points are listed in \cref{tab:upper_limits_offdiag_d3}. 
\\
The strongest limit is obtained on the $\mathring{a}_{\mu\tau}^{(3)}$ coefficient, which is plausible, since this coefficient introduces oscillations between the $\nu_\mu$ and $\nu_\tau$ flavour. As discussed in \cref{sec:data_sample_and_selection}, most events in the data sample are track-like events resulting from $\nu_\mu$ CC interactions, which make up most of the HPT and LPT event classes. On the other hand, most events caused by $\nu_\tau$ CC interactions result in shower-like events, which are primarily located in the Showers event class. Hence, the relatively strong sensitivity to $\mathring{a}_{\mu\tau}^{(3)}$ can be explained by the significant track disappearance and the mild shower appearance resulting from a nonzero value of this coefficient. A similar effect could be expected from $\mathring{a}_{e\mu}^{(3)}$. However, oscillations in and from the $\nu_e$ flavour are suppressed because of the influence of the matter potential $V_{e}$, which sets the $\nu_e$ flavour on a higher potential, explaining the comparably low sensitivity to this coefficient. In contrast, the $\mathring{a}_{e\tau}^{(3)}$ coefficient induces oscillations between flavours that primarily yield shower-like events, to which this data sample is less sensitive.
\begin{table}[!h]
  \centering
  \small
  \caption{Best fits and upper limits for the isotropic LIV coefficients
  $\mathring{a}_{\mu\tau}^{(3)}$, $\mathring{a}_{e\mu}^{(3)}$, and $\mathring{a}_{e\tau}^{(3)}$.}
  \renewcommand{\arraystretch}{1.8}
  {%
  \setlength{\tabcolsep}{14pt}
  \begin{tabular}{@{} C c c c c @{}}
    \toprule
    \multicolumn{1}{c}{\textbf{LIV coefficient}} &
    \multicolumn{1}{c}{\textbf{Unit}} &
    \multicolumn{1}{c}{\textbf{Best fit}} &
    \multicolumn{1}{c}{\textbf{95\% CL}} &
    \multicolumn{1}{c}{\textbf{99\% CL}} \\
    \midrule
    \mathring{a}_{\mu\tau}^{(3)} & $10^{-23}\,\si{GeV}$ & $9.44 \times 10^{-2}$ & 0.48 & 0.57 \\
    \mathring{a}_{e \mu}^{(3)} & $10^{-23}\,\si{GeV}$ & $4.97 \times 10^{-5}$ & 2.53 & 3.02 \\
    \mathring{a}_{e\tau}^{(3)}  & $10^{-23}\,\si{GeV}$ & $4.22 \times 10^{-10}$ & 4.70 & 5.75 \\
    \bottomrule
  \end{tabular}%
  }
  \label{tab:upper_limits_offdiag_d3}
\end{table}

\subsection{Diagonal coefficients}

\begin{figure}[!h]
\centering

% Row 1
\begin{subfigure}{0.49\textwidth}
  \includegraphics[width=\linewidth]{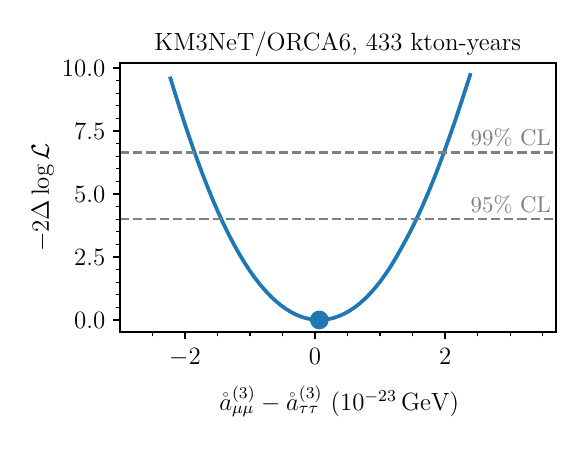}
  \caption{$d=3$}
  \label{fig:diag_md3}
\end{subfigure}\hfill
\begin{subfigure}{0.49\textwidth}
  \includegraphics[width=\linewidth]{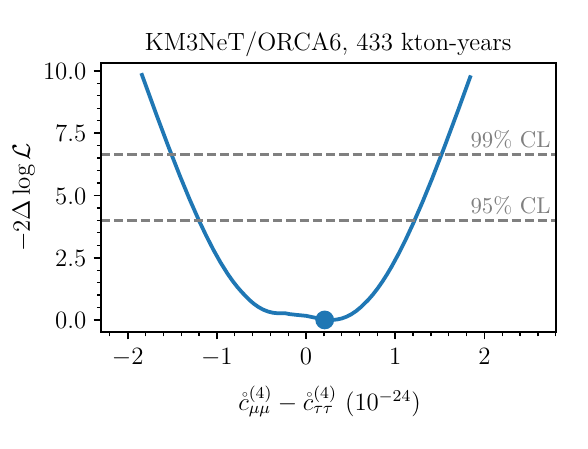}
  \caption{$d=4$}
  \label{fig:diag_md4}
\end{subfigure}

\vspace{0.5em}

% Row 2
\begin{subfigure}{0.49\textwidth}
  \includegraphics[width=\linewidth]{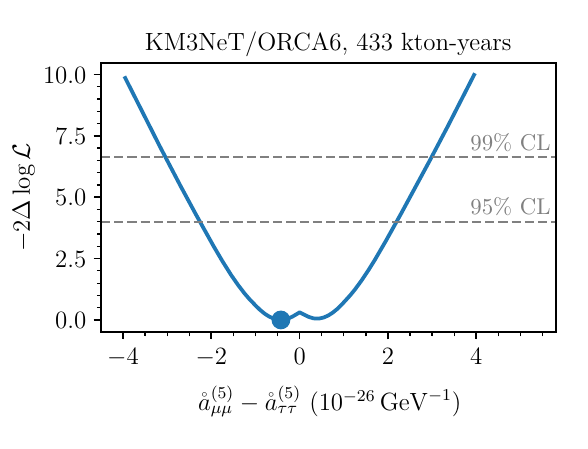}
  \caption{$d=5$}
  \label{fig:diag_md5}
\end{subfigure}\hfill
\begin{subfigure}{0.49\textwidth}
  \includegraphics[width=\linewidth]{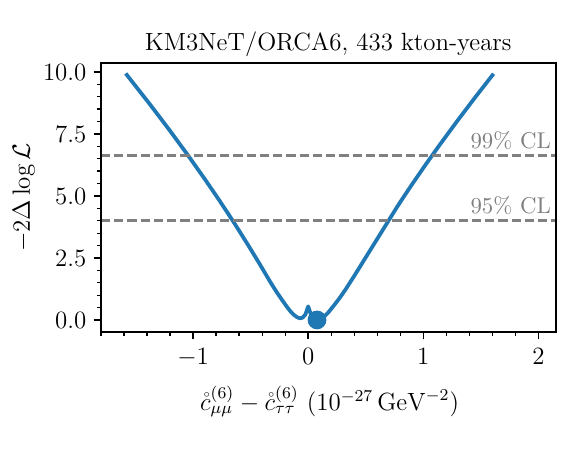}
  \caption{$d=6$}
  \label{fig:diag_md6}
\end{subfigure}

\vspace{0.5em}

% Row 3
\begin{subfigure}{0.49\textwidth}
  \includegraphics[width=\linewidth]{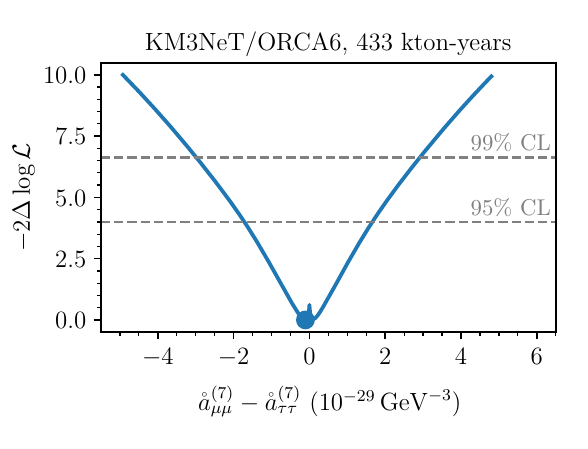}
  \caption{$d=7$}
  \label{fig:diag_md7}
\end{subfigure}\hfill
\begin{subfigure}{0.49\textwidth}
  \includegraphics[width=\linewidth]{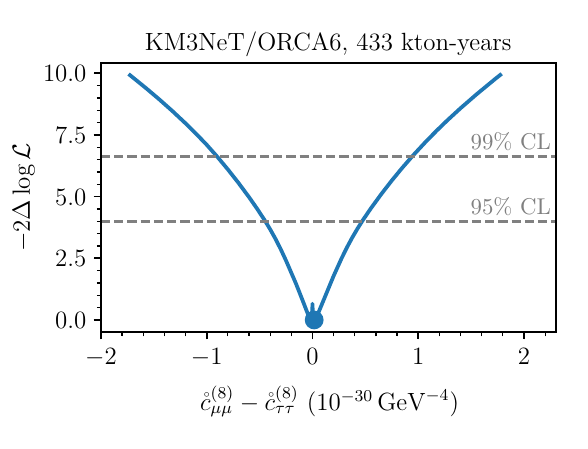}
  \caption{$d=8$}
  \label{fig:diag_md8}
\end{subfigure}
\caption{Likelihood-ratio scans for diagonal coefficient combinations across $d = 3, \dots, 8$. The blue dot in each panel indicates the respective best-fit point for that dimension. The intersections with the dashed lines define the upper limits.}
\label{fig:diagonal_md3to8_lr_scans}
\end{figure}

The upper limits on the diagonal coefficients are shown in \cref{fig:diagonal_md3to8_lr_scans}. They are listed, together with the corresponding best-fit points, in \cref{tab:upper_limits_diag}. 
\\
It is noteworthy that the shapes of the likelihood ratio curves are nearly but not completely symmetric around zero for these coefficients. This behaviour arises from the distinct effects that the sign of LIV coefficients has on neutrinos and antineutrinos. Since in the analysis no distinction is made between neutrinos and antineutrinos in the data sample, a perfect symmetry might be expected. However, the atmospheric neutrino flux comprises unequal fractions of neutrinos and antineutrinos, and their interaction cross sections differ. Consequently, interchanging the oscillation probabilities between neutrinos and antineutrinos induces a slight asymmetry in the observed distributions. Therefore, the symmetry seen in the likelihood ratios is only approximate, rather than exact.

\begin{table}[!h]
  \centering
  \small % a touch smaller than \small
  \caption{Best fits and upper limits for the diagonal LIV differences
  $\mathring{k}_{\mu\mu}^{(d)} - \mathring{k}_{\tau\tau}^{(d)}$ for $d=3,\ldots,8$.}
  \renewcommand{\arraystretch}{1.8}
  {%
  \setlength{\tabcolsep}{14pt} % was 12pt; keep this small for 7 columns
  \begin{tabular}{@{} C c S *{2}{S} *{2}{S} @{}}
    \toprule
    \multicolumn{1}{c}{\textbf{LIV coefficient}} &
    \multicolumn{1}{c}{\textbf{Unit}} &
    \multicolumn{1}{c}{\textbf{Best fit}} &
    \multicolumn{2}{c}{\textbf{95\% CL}} &
    \multicolumn{2}{c}{\textbf{99\% CL}} \\
    \cmidrule(lr){4-5}\cmidrule(lr){6-7}
    & & & \multicolumn{1}{c}{\textbf{low}} & \multicolumn{1}{c}{\textbf{high}}
          & \multicolumn{1}{c}{\textbf{low}} & \multicolumn{1}{c}{\textbf{high}} \\
    \midrule
    \mathring{a}_{\mu\mu}^{(3)} - \mathring{a}_{\tau\tau}^{(3)} & $10^{-23}\,\si{GeV}$      & \multicolumn{1}{c}{0.07} & \multicolumn{1}{c}{-1.44} & \multicolumn{1}{c}{1.56} & \multicolumn{1}{c}{-1.85} & \multicolumn{1}{c}{1.98} \\
    \mathring{c}_{\mu\mu}^{(4)} - \mathring{c}_{\tau\tau}^{(4)} & $10^{-24}$             & \multicolumn{1}{c}{0.28} & \multicolumn{1}{c}{-1.20} & \multicolumn{1}{c}{1.22} & \multicolumn{1}{c}{-1.51} & \multicolumn{1}{c}{1.52} \\
    \mathring{a}_{\mu\mu}^{(5)} - \mathring{a}_{\tau\tau}^{(5)} & $10^{-26}\,\si{GeV^{-1}}$ & \multicolumn{1}{c}{-0.48} & \multicolumn{1}{c}{-2.26} & \multicolumn{1}{c}{2.20} & \multicolumn{1}{c}{-3.04} & \multicolumn{1}{c}{2.99} \\
    \mathring{c}_{\mu\mu}^{(6)} - \mathring{c}_{\tau\tau}^{(6)} & $10^{-27}\,\si{GeV^{-2}}$ & \multicolumn{1}{c}{0.08} & \multicolumn{1}{c}{-0.65} & \multicolumn{1}{c}{0.69} & \multicolumn{1}{c}{-1.04} & \multicolumn{1}{c}{1.08} \\
    \mathring{a}_{\mu\mu}^{(7)} - \mathring{a}_{\tau\tau}^{(7)} & $10^{-29}\,\si{GeV^{-3}}$ & \multicolumn{1}{c}{-0.10} & \multicolumn{1}{c}{-1.72} & \multicolumn{1}{c}{1.66} & \multicolumn{1}{c}{-2.99} & \multicolumn{1}{c}{2.92} \\
    \mathring{c}_{\mu\mu}^{(8)} - \mathring{c}_{\tau\tau}^{(8)} & $10^{-31}\,\si{GeV^{-4}}$ & \multicolumn{1}{c}{0.16} & \multicolumn{1}{c}{-4.46} & \multicolumn{1}{c}{4.69} & \multicolumn{1}{c}{-9.04} & \multicolumn{1}{c}{9.44}\\
    \bottomrule
  \end{tabular}%
  }
  \label{tab:upper_limits_diag}
\end{table}

\subsection{Comparison with other experiments}
\begin{figure}[!h]
\centering
  \includegraphics[width=\linewidth]{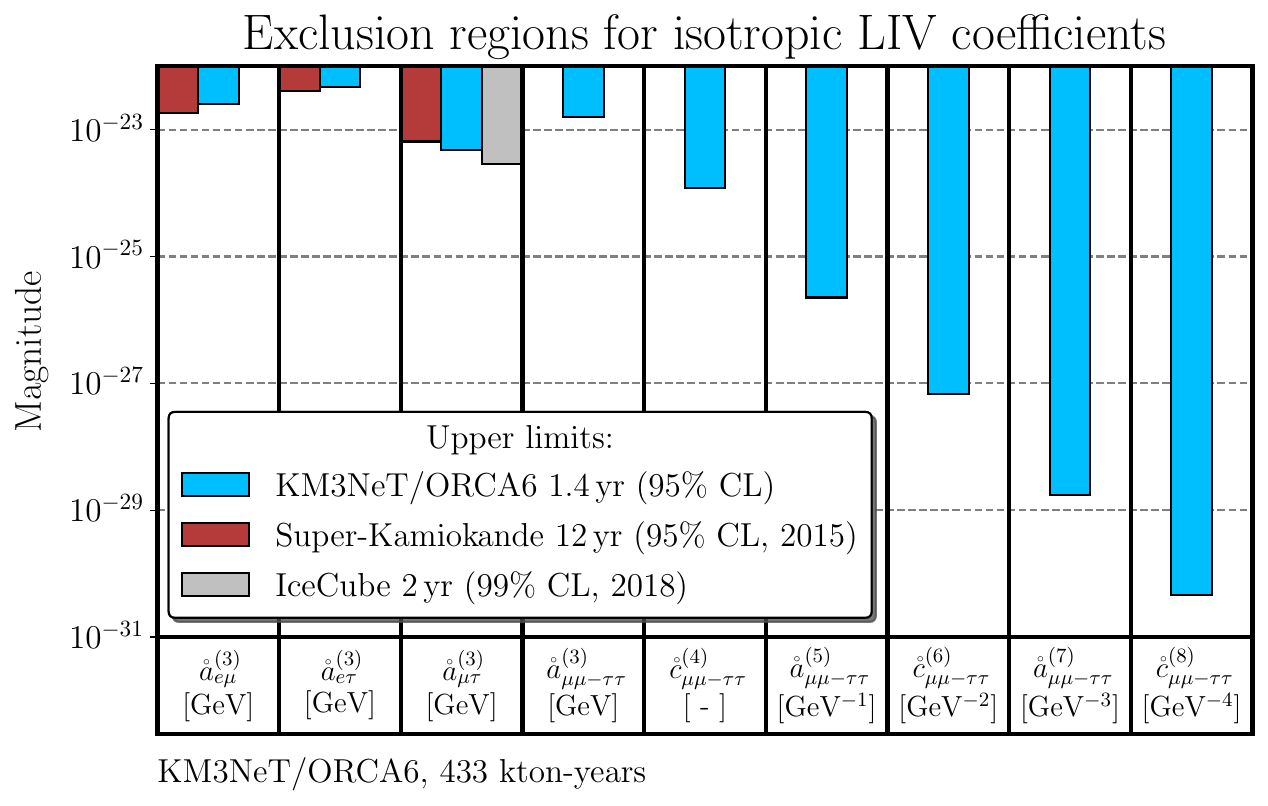}
  \caption{Exclusion regions for isotropic LIV coefficients obtained with 1.4 years of ORCA6 data. The regions are compared to previous upper limits set with 12 years of Super-Kamiokande data \cite{SK_paper} and 2 years of IceCube data \cite{IC2yrLIV}.}
  \label{fig:barplot}
\end{figure}

In \cref{fig:barplot} the results of this analysis are compared with those from approximately 12 years of Super-Kamiokande data \cite{SK_paper}, as well as a two-year IceCube analysis \cite{IC2yrLIV} based on atmospheric neutrinos. This analysis provides competitive results despite the small detector size and shorter livetime of ORCA6. An additional IceCube analysis employing high-energy cosmic neutrinos \cite{ic_astro} reported results which are strongly dependent on the assumed initial flavour ratio at the source. The corresponding IceCube results are not included in \cref{fig:barplot}, since for many plausible flavour ratios, no upper limit could be established.
\\
ORCA6 provides the first upper limits on isotropic diagonal LIV coefficients in the neutrino sector that do not rely on assumptions about the initial astronomical flavour ratios. The model-independent approach enhances the robustness and general applicability of the constraints obtained by this analysis.

%% file: sections/conclusion.tex
\section{Conclusion}

A search for isotropic Lorentz invariance violation using 1.4 years of atmospheric neutrino data collected with the first six detection units of the KM3NeT/ORCA detector has been conducted. No evidence for LIV was observed; upper limits were therefore set on a subset of isotropic LIV coefficients. These bounds not only complement the existing constraints set by experiments such as Super-Kamiokande and IceCube, but also provide new upper limits on several, previously unconstrained coefficients.
\\
This analysis provides the first experimental constraints on diagonal LIV coefficients that do not rely on assumptions about the initial astronomical neutrino flavour ratios. This model-independent approach enhances the robustness and generality of the constraints.
\\
The present work lays the foundation for future analyses with larger data samples and an extended detector configuration, which are expected to improve the potential to explore subleading effects and probe additional LIV scenarios.

%% file: sections/acknowledgments.tex
A. Domi acknowledges the support from the European Union’s Horizon 2021 research and innovation programme under the Marie Skłodowska-Curie grant agreement No. 101068013 (QGRANT), and from the Emerging Talent Initiative funding at the Friedrich-Alexander-Universität Erlangen-Nürnberg. The authors acknowledge the financial support of:
%INFRADEV
KM3NeT-INFRADEV2 project, funded by the European Union Horizon Europe Research and Innovation Programme under grant agreement No 101079679;
%Belgium
Funds for Scientific Research (FRS-FNRS), Francqui foundation, BAEF foundation.
%Czeck
Czech Science Foundation (GAČR 24-12702S);
%France
Agence Nationale de la Recherche (contract ANR-15-CE31-0020), Centre National de la Recherche Scientifique (CNRS), Commission Europ\'eenne (FEDER fund and Marie Curie Program), LabEx UnivEarthS (ANR-10-LABX-0023 and ANR-18-IDEX-0001), Paris \^Ile-de-France Region, Normandy Region (Alpha, Blue-waves and Neptune), France,
%For the CPER
The Provence-Alpes-Côte d'Azur Delegation for Research and Innovation (DRARI), the Provence-Alpes-Côte d'Azur region, the Bouches-du-Rhône Departmental Council, the Metropolis of Aix-Marseille Provence and the City of Marseille through the CPER 2021-2027 NEUMED project,
%For IN2P3
The CNRS Institut National de Physique Nucléaire et de Physique des Particules (IN2P3);
%Georgia
Shota Rustaveli National Science Foundation of Georgia (SRNSFG, FR-22-13708), Georgia;
%Germany (Max Planck Inst.)
This research was funded by the European Union (ERC MuSES project No 101142396);
%Greece
The General Secretariat of Research and Innovation (GSRI), Greece;
%Italy
Istituto Nazionale di Fisica Nucleare (INFN) and Ministero dell’Universit{\`a} e della Ricerca (MUR), through PRIN 2022 program (Grant PANTHEON 2022E2J4RK, Next Generation EU) and PON R\&I program (Avviso n. 424 del 28 febbraio 2018, Progetto PACK-PIR01 00021), Italy; IDMAR project Po-Fesr Sicilian Region az. 1.5.1; A. De Benedittis, W. Idrissi Ibnsalih, M. Bendahman, A. Nayerhoda, G. Papalashvili, I. C. Rea, A. Simonelli have been supported by the Italian Ministero dell'Universit{\`a} e della Ricerca (MUR), Progetto CIR01 00021 (Avviso n. 2595 del 24 dicembre 2019); KM3NeT4RR MUR Project National Recovery and Resilience Plan (NRRP), Mission 4 Component 2 Investment 3.1, Funded by the European Union – NextGenerationEU,CUP I57G21000040001, Concession Decree MUR No. n. Prot. 123 del 21/06/2022;
%Morocco
Ministry of Higher Education, Scientific Research and Innovation, Morocco, and the Arab Fund for Economic and Social Development, Kuwait;
%The Netherlands
Nederlandse organisatie voor Wetenschappelijk Onderzoek (NWO), the Netherlands;
%Poland
The grant “AstroCeNT: Particle Astrophysics Science and Technology Centre”, carried out within the International Research Agendas programme of the Foundation for Polish Science financed by the European Union under the European Regional Development Fund; The program: “Excellence initiative-research university” for the AGH University in Krakow; The ARTIQ project: UMO-2021/01/2/ST6/00004 and ARTIQ/0004/2021;
%Romania
Ministry of Education and Scientific Research, Romania
%Slovak Republic
Slovak Research and Development Agency under Contract No. APVV-22-0413; Ministry of Education, Research, Development and Youth of the Slovak Republic;
%Spain
MCIN for PID2021-124591NB-C41, -C42, -C43 and PDC2023-145913-I00 funded by MCIN/AEI/10.13039/501100011033 and by “ERDF A way of making Europe”, for ASFAE/2022/014 and ASFAE/2022 /023 with funding from the EU NextGenerationEU (PRTR-C17.I01) and Generalitat Valenciana, for Grant AST22\_6.2 with funding from Consejer\'{\i}a de Universidad, Investigaci\'on e Innovaci\'on and Gobierno de Espa\~na and European Union - NextGenerationEU, for CSIC-INFRA23013 and for CNS2023-144099, Generalitat Valenciana for CIDEGENT/2020/049, CIDEGENT/2021/23, CIDEIG/2023/20, ESGENT2024/24, CIPROM/2023/51, GRISOLIAP/2021/192 and INNVA1/2024/110 (IVACE+i), Spain;
%UAE
Khalifa University internal grants (ESIG-2023-008, RIG-2023-070 and RIG-2024-047), United Arab Emirates;
%UK
The European Union's Horizon 2020 Research and Innovation Programme (ChETEC-INFRA - Project no. 101008324).
% disclaimer
Views and opinions expressed are those of the author(s) only and do not necessarily reflect those of the European Union or the European Research Council. Neither the European Union nor the granting authority can be held responsible for them.

%% file: sections/author_list.tex
\clearpage
\section*{The KM3NeT collaboration}

\begingroup
\small
\setlength{\parindent}{0pt}
\setlength{\parskip}{0pt}

% Make ORCID optional
\providecommand{\orcidlink}[1]{}

% Ignore Elsevier-only commands if present in DB output
\providecommand{\cortext}[2][]{}
\providecommand{\ead}[1]{}

% Corresponding author marker
\providecommand{\corref}[1]{\textsuperscript{*}}

% Local author/address commands (avoid redefining \author)
\newcommand{\collabauthor}[2][]{#2\textsuperscript{#1}, }
\newcommand{\address}[2][]{\par\textsuperscript{#1}\, #2}

\cortext[cor]{corresponding author}

\collabauthor[b,a]{O.~Adriani\,\orcidlink{0000-0002-3592-0654}}
\collabauthor[c,bf]{A.~Albert}
\collabauthor[d]{A.\,R.~Alhebsi\,\orcidlink{0009-0002-7320-7638}}
\collabauthor[d]{S.~Alshalloudi}
\collabauthor[e]{S. Alves Garre\,\orcidlink{0000-0003-1893-0858}}
\collabauthor[f]{F.~Ameli}
\collabauthor[g]{M.~Andre}
\collabauthor[h]{L.~Aphecetche\,\orcidlink{0000-0001-7662-3878}}
\collabauthor[i]{M. Ardid\,\orcidlink{0000-0002-3199-594X}}
\collabauthor[i]{S. Ardid\,\orcidlink{0000-0003-4821-6655}}
\collabauthor[j]{J.~Aublin}
\collabauthor[l,k]{F.~Badaracco\,\orcidlink{0000-0001-8553-7904}}
\collabauthor[m]{L.~Bailly-Salins}
\collabauthor[j]{B.~Baret}
\collabauthor[e]{A. Bariego-Quintana\,\orcidlink{0000-0001-5187-7505}}
\collabauthor[k,l]{L.~Barigione}
\collabauthor[n]{M.~Barnard\,\orcidlink{0000-0003-1720-7959}}
\collabauthor[j]{Y.~Becherini}
\collabauthor[o]{M.~Bendahman}
\collabauthor[q,p]{F.~Benfenati~Gualandi}
\collabauthor[r,o]{M.~Benhassi}
\collabauthor[s]{D.\,M.~Benoit\,\orcidlink{0000-0002-7773-6863}}
\collabauthor[u,t]{Z. Be\v{n}u\v{s}ov\'a\,\orcidlink{0000-0002-2677-7657}}
\collabauthor[v]{E.~Berbee}
\collabauthor[v]{C.~van~Bergen}
\collabauthor[b]{E.~Berti}
\collabauthor[w]{V.~Bertin\,\orcidlink{0000-0001-6688-4580}}
\collabauthor[b]{P.~Betti\,\orcidlink{0000-0002-7097-165X}}
\collabauthor[x]{S.~Biagi\,\orcidlink{0000-0001-8598-0017}}
\collabauthor[n]{M.~Boettcher}
\collabauthor[x]{D.~Bonanno\,\orcidlink{0000-0003-0223-3580}}
\collabauthor[y]{M.~Bond{\`\i}}
\collabauthor[a,b]{M.~Bongi\,\orcidlink{0000-0002-6050-1937}}
\collabauthor[b]{S.~Bottai}
\collabauthor[bg]{A.\,B.~Bouasla}
\collabauthor[z]{J.~Boumaaza}
\collabauthor[w]{M.~Bouta}
\collabauthor[aa,o]{C.~Bozza\,\orcidlink{0009-0006-3741-2676}}
\collabauthor[ab,o]{R.\,M.~Bozza}
\collabauthor[ac]{H.~Br\^{a}nza\c{s}}
\collabauthor[h]{F.~Bretaudeau}
\collabauthor[w]{M.~Breuhaus\,\orcidlink{0000-0003-0268-5122}}
\collabauthor[ad,v]{R.~Bruijn}
\collabauthor[w]{J.~Brunner}
\collabauthor[y]{R.~Bruno\,\orcidlink{0000-0002-3517-6597}}
\collabauthor[v]{E.~Buis}
\collabauthor[r,o]{R.~Buompane}
\collabauthor[e]{I.~Burriel}
\collabauthor[w]{J.~Busto}
\collabauthor[l]{B.~Caiffi}
\collabauthor[e]{D.~Calvo}
\collabauthor[v]{E.G.J.~van~Campenhout}
\collabauthor[f,ae]{A.~Capone}
\collabauthor[q,p]{F.~Carenini}
\collabauthor[ad,v]{V.~Carretero\,\orcidlink{0000-0002-7540-0266}}
\collabauthor[j]{T.~Cartraud}
\collabauthor[af,p]{P.~Castaldi}
\collabauthor[e]{V.~Cecchini\,\orcidlink{0000-0003-4497-2584}}
\collabauthor[f,ae]{S.~Celli}
\collabauthor[ag]{M.~Chabab}
\collabauthor[ah]{A.~Chen\,\orcidlink{0000-0001-6425-5692}}
\collabauthor[ai,x]{S.~Cherubini}
\collabauthor[p]{T.~Chiarusi}
\collabauthor[aj]{W.~Chung}
\collabauthor[ak]{M.~Circella\,\orcidlink{0000-0002-5560-0762}}
\collabauthor[al]{R.~Clark}
\collabauthor[x]{R.~Cocimano}
\collabauthor[j]{J.\,A.\,B.~Coelho}
\collabauthor[j]{A.~Coleiro}
\collabauthor[j]{A. Condorelli}
\collabauthor[x]{R.~Coniglione\,\orcidlink{0000-0002-8289-5447}}
\collabauthor[w]{P.~Coyle}
\collabauthor[j]{A.~Creusot}
\collabauthor[x]{G.~Cuttone}
\collabauthor[h]{R.~Dallier\,\orcidlink{0000-0001-9452-4849}}
\collabauthor[r,o]{A.~De~Benedittis}
\collabauthor[h]{X. de La Bernardie\,\orcidlink{0000-0001-8288-9787}}
\collabauthor[al]{G.~De~Wasseige\,\orcidlink{0000-0002-1010-5100}}
\collabauthor[h]{V.~Decoene}
\collabauthor[w]{P. Deguire}
\collabauthor[q,p]{I.~Del~Rosso}
\collabauthor[x]{L.\,S.~Di~Mauro}
\collabauthor[f,ae]{I.~Di~Palma\,\orcidlink{0000-0003-1544-8943}}
\collabauthor[am]{A.\,F.~D\'\i{}az\,\orcidlink{0000-0002-2615-6586}}
\collabauthor[bh,x]{D.~Diego-Tortosa\,\orcidlink{0000-0001-5546-3748}}
\collabauthor[x]{C.~Distefano\,\orcidlink{0000-0001-8632-1136}}
\collabauthor[an]{A.~Domi\,\orcidlink{0000-0002-9610-7296}\corref{cor}}
\ead{km3net-pc@km3net.de; adomi@km3net.de}
\collabauthor[j]{C.~Donzaud}
\collabauthor[w]{D.~Dornic\,\orcidlink{0000-0001-5729-1468}}
\collabauthor[ao]{E.~Drakopoulou\,\orcidlink{0000-0003-2493-8039}}
\collabauthor[c,bf]{D.~Drouhin\,\orcidlink{0000-0002-9719-2277}}
\collabauthor[w]{J.-G. Ducoin}
\collabauthor[j]{P.~Duverne}
\collabauthor[u]{R. Dvornick\'{y}\,\orcidlink{0000-0002-4401-1188}}
\collabauthor[an]{T.~Eberl\,\orcidlink{0000-0002-5301-9106}}
\collabauthor[u,t]{E. Eckerov\'{a}\,\orcidlink{0000-0001-9438-724X}}
\collabauthor[z]{A.~Eddymaoui}
\collabauthor[j]{M.~Eff}
\collabauthor[v]{D.~van~Eijk}
\collabauthor[ap]{I.~El~Bojaddaini}
\collabauthor[j]{S.~El~Hedri}
\collabauthor[w]{S.~El~Mentawi}
\collabauthor[l]{V.~Ellajosyula}
\collabauthor[w]{A.~Enzenh\"ofer}
\collabauthor[aj]{M.~Farino\,\orcidlink{0000-0002-1649-3618}}
\collabauthor[aq,o]{A.~Ferrara}
\collabauthor[ai,x]{G.~Ferrara}
\collabauthor[ar]{M.~D.~Filipovi\'c\,\orcidlink{0000-0002-4990-9288}}
\collabauthor[p]{F.~Filippini}
\collabauthor[w]{A.~Foisseau\,\orcidlink{0009-0007-9457-4599}}
\collabauthor[x]{D.~Franciotti}
\collabauthor[aa,o]{L.\,A.~Fusco\,\orcidlink{0000-0001-8254-3372}}
\collabauthor[ae,f]{S.~Gagliardini}
\collabauthor[an]{T.~Gal\,\orcidlink{0000-0001-7821-8673}}
\collabauthor[i]{J.~Garc{\'\i}a~M{\'e}ndez\,\orcidlink{0000-0002-1580-0647}}
\collabauthor[e]{A.~Garcia~Soto\,\orcidlink{0000-0002-8186-2459}}
\collabauthor[v]{C.~Gatius~Oliver\,\orcidlink{0009-0002-1584-1788}}
\collabauthor[an]{N.~Gei{\ss}elbrecht}
\collabauthor[ap]{H.~Ghaddari}
\collabauthor[r,o]{L.~Gialanella}
\collabauthor[s]{B.\,K.~Gibson}
\collabauthor[x]{E.~Giorgio}
\collabauthor[j]{I.~Goos\,\orcidlink{0009-0008-1479-539X}}
\collabauthor[j]{P.~Goswami}
\collabauthor[e]{S.\,R.~Gozzini\,\orcidlink{0000-0001-5152-9631}}
\collabauthor[an]{R.~Gracia}
\collabauthor[m]{B.~Guillon}
\collabauthor[an]{C.~Haack}
\collabauthor[aj]{C.~Hanna\,\orcidlink{0000-0003-4764-1270}}
\collabauthor[as]{H.~van~Haren}
\collabauthor[aj]{E.~Hazelton}
\collabauthor[v]{A.~Heijboer}
\collabauthor[an]{L.~Hennig\,\orcidlink{0000-0002-2816-2242}\corref{cor}}
\ead{lukas.hennig@fau.de}
\collabauthor[e]{J.\,J.~Hern{\'a}ndez-Rey}
\collabauthor[x]{A.~Idrissi\,\orcidlink{0000-0001-8936-6364}}
\collabauthor[o]{W.~Idrissi~Ibnsalih}
\collabauthor[p]{G.~Illuminati}
\collabauthor[e]{R.~Jaimes}
\collabauthor[an]{O.~Janik\,\orcidlink{0009-0007-3121-2486}}
\collabauthor[w]{D.~Joly}
\collabauthor[at,v]{M.~de~Jong}
\collabauthor[ad,v]{P.~de~Jong}
\collabauthor[v]{B.\,J.~Jung}
\collabauthor[bi,au]{P.~Kalaczy\'nski\,\orcidlink{0000-0001-9278-5906}}
\collabauthor[av]{T.~Kapoor\,\orcidlink{0000-0001-5726-3037}}
\collabauthor[an]{U.\,F.~Katz}
\collabauthor[s]{J.~Keegans}
\collabauthor[aw]{T.~Khvichia}
\collabauthor[ax,aw]{G.~Kistauri}
\collabauthor[an]{C.~Kopper\,\orcidlink{0000-0001-6288-7637}}
\collabauthor[ay,j]{A.~Kouchner}
\collabauthor[az]{Y. Y. Kovalev\,\orcidlink{0000-0001-9303-3263}}
\collabauthor[t]{L.~Krupa}
\collabauthor[v]{V.~Kueviakoe}
\collabauthor[l]{V.~Kulikovskiy\,\orcidlink{0000-0003-4096-5934}}
\collabauthor[ax]{R.~Kvatadze}
\collabauthor[m]{M.~Labalme}
\collabauthor[an]{R.~Lahmann}
\collabauthor[j]{M.~Lamoureux\,\orcidlink{0000-0002-8860-5826}}
\collabauthor[aj]{A.~Langella\,\orcidlink{0000-0001-6273-3558}}
\collabauthor[x]{G.~Larosa}
\collabauthor[m]{C.~Lastoria}
\collabauthor[al]{J.~Lazar}
\collabauthor[e]{A.~Lazo}
\collabauthor[m]{G.~Lehaut}
\collabauthor[al]{V.~Lema{\^\i}tre}
\collabauthor[y]{E.~Leonora}
\collabauthor[e]{N.~Lessing\,\orcidlink{0000-0001-8670-2780}}
\collabauthor[q,p]{G.~Levi\,\orcidlink{0000-0003-1714-6359}}
\collabauthor[j]{M.~Lindsey~Clark}
\collabauthor[y]{F.~Longhitano}
\collabauthor[j]{M.~Loup}
\collabauthor[n]{A.~Luashvili\,\orcidlink{0000-0003-4384-1638}}
\collabauthor[e]{S.~Madarapu}
\collabauthor[w]{F.~Magnani}
\collabauthor[l,k]{L.~Malerba}
\collabauthor[t]{F.~Mamedov}
\collabauthor[o]{A.~Manfreda\,\orcidlink{0000-0002-0998-4953}}
\collabauthor[ba]{A.~Manousakis}
\collabauthor[k,l]{M.~Marconi\,\orcidlink{0009-0008-0023-4647}}
\collabauthor[q,p]{A.~Margiotta\,\orcidlink{0000-0001-6929-5386}}
\collabauthor[ab,o]{A.~Marinelli}
\collabauthor[ao]{C.~Markou}
\collabauthor[h]{L.~Martin\,\orcidlink{0000-0002-9781-2632}}
\collabauthor[ae,f]{M.~Mastrodicasa}
\collabauthor[o]{S.~Mastroianni\,\orcidlink{0000-0002-9467-0851}}
\collabauthor[al]{J.~Mauro\,\orcidlink{0009-0005-9324-7970}}
\collabauthor[au]{K.\,C.\,K.~Mehta\,\orcidlink{0009-0005-2831-6917}}
\collabauthor[ab,o]{G.~Miele}
\collabauthor[o]{P.~Migliozzi\,\orcidlink{0000-0001-5497-3594}}
\collabauthor[x]{E.~Migneco}
\collabauthor[r,o]{M.\,L.~Mitsou}
\collabauthor[o]{C.\,M.~Mollo\,\orcidlink{0000-0003-2766-8003}}
\collabauthor[r,o]{L. Morales-Gallegos\,\orcidlink{0000-0002-2241-4365}}
\collabauthor[b]{N.~Mori\,\orcidlink{0000-0003-2138-3787}}
\collabauthor[ap]{A.~Moussa\,\orcidlink{0000-0003-2233-9120}}
\collabauthor[m]{I.~Mozun~Mateo}
\collabauthor[p]{R.~Muller\,\orcidlink{0000-0002-5247-7084}}
\collabauthor[r,o]{M.\,R.~Musone}
\collabauthor[x]{M.~Musumeci\,\orcidlink{0000-0002-9384-4805}}
\collabauthor[bb]{S.~Navas\,\orcidlink{0000-0003-1688-5758}}
\collabauthor[ak]{A.~Nayerhoda}
\collabauthor[f]{C.\,A.~Nicolau}
\collabauthor[ah]{B.~Nkosi\,\orcidlink{0000-0003-0954-4779}}
\collabauthor[l]{B.~{\'O}~Fearraigh\,\orcidlink{0000-0002-1795-1617}}
\collabauthor[ab,o]{V.~Oliviero\,\orcidlink{0009-0004-9638-0825}}
\collabauthor[x]{A.~Orlando}
\collabauthor[j]{E.~Oukacha}
\collabauthor[a,b]{L.~Pacini\,\orcidlink{0000-0001-6808-9396}}
\collabauthor[x]{D.~Paesani}
\collabauthor[e]{J.~Palacios~Gonz{\'a}lez\,\orcidlink{0000-0001-9292-9981}}
\collabauthor[ak,aw]{G.~Papalashvili\,\orcidlink{0000-0002-4388-2643}}
\collabauthor[b]{P.~Papini}
\collabauthor[k,l]{V.~Parisi}
\collabauthor[m]{A.~Parmar\,\orcidlink{0009-0006-7193-8524}}
\collabauthor[e]{G.~Pascua}
\collabauthor[i]{B. Pascual-Estrugo\,\orcidlink{0009-0002-9109-5799}}
\collabauthor[ak]{C.~Pastore}
\collabauthor[ac]{A.~M.~P{\u a}un}
\collabauthor[ac]{G.\,E.~P\u{a}v\u{a}la\c{s}}
\collabauthor[j]{S. Pe\~{n}a Mart\'inez\,\orcidlink{0000-0001-8939-0639}}
\collabauthor[w]{M.~Perrin-Terrin}
\collabauthor[m]{V.~Pestel}
\collabauthor[t,bj]{M.~Petropavlova\,\orcidlink{0000-0002-0416-0795}}
\collabauthor[x]{P.~Piattelli}
\collabauthor[az,bk]{A.~Plavin}
\collabauthor[aa,o]{C.~Poir{\`e}}
\collabauthor[c]{T.~Pradier\,\orcidlink{0000-0001-5501-0060}}
\collabauthor[e]{J.~Prado}
\collabauthor[x]{S.~Pulvirenti\,\orcidlink{0000-0003-3017-512X}}
\collabauthor[y]{N.~Randazzo}
\collabauthor[bc]{A.~Ratnani}
\collabauthor[bd]{S.~Razzaque}
\collabauthor[o]{I.\,C.~Rea\,\orcidlink{0000-0002-3954-7754}}
\collabauthor[e]{D.~Real\,\orcidlink{0000-0002-1038-7021}}
\collabauthor[x]{G.~Riccobene\,\orcidlink{0000-0002-0600-2774}}
\collabauthor[n]{J.~Robinson}
\collabauthor[m]{A.~Romanov}
\collabauthor[az]{E.~Ros\,\orcidlink{0000-0001-9503-4892}}
\collabauthor[e]{A. \v{S}aina}
\collabauthor[e]{F.~Salesa~Greus\,\orcidlink{0000-0002-8610-8703}}
\collabauthor[at,v]{D.\,F.\,E.~Samtleben}
\collabauthor[e]{A.~S{\'a}nchez~Losa\,\orcidlink{0000-0001-9596-7078}}
\collabauthor[x]{S.~Sanfilippo}
\collabauthor[k,l]{M.~Sanguineti}
\collabauthor[x]{D.~Santonocito}
\collabauthor[x]{P.~Sapienza}
\collabauthor[b]{M.~Scaringella}
\collabauthor[al,j]{M.~Scarnera}
\collabauthor[an]{J.~Schnabel}
\collabauthor[an]{J.~Schumann\,\orcidlink{0000-0003-3722-086X}}
\collabauthor[d]{M.~Senniappan\,\orcidlink{0000-0001-6734-7699}}
\collabauthor[al]{P. A.~Sevle~Myhr\,\orcidlink{0009-0005-9103-4410}}
\collabauthor[ak]{I.~Sgura}
\collabauthor[aw]{R.~Shanidze}
\collabauthor[bl,w]{Chengyu Shao\,\orcidlink{0000-0002-2954-1180}}
\collabauthor[j]{A.~Sharma}
\collabauthor[t]{Y.~Shitov}
\collabauthor[u]{F. \v{S}imkovic}
\collabauthor[o]{A.~Simonelli}
\collabauthor[y]{A.~Sinopoulou\,\orcidlink{0000-0001-9205-8813}}
\collabauthor[w]{C.~Sironneau\,\orcidlink{0000-0003-3762-635X}}
\collabauthor[o]{B.~Spisso}
\collabauthor[q,p]{M.~Spurio\,\orcidlink{0000-0002-8698-3655}}
\collabauthor[b]{O.~Starodubtsev}
\collabauthor[t]{I. \v{S}tekl}
\collabauthor[h]{D.~Stocco\,\orcidlink{0000-0002-5377-5163}}
\collabauthor[k,l]{M.~Taiuti}
\collabauthor[z,bc]{Y.~Tayalati}
\collabauthor[e]{J.~Tena\,\orcidlink{0000-0002-1300-6781}}
\collabauthor[n]{H.~Thiersen}
\collabauthor[d]{S.~Thoudam}
\collabauthor[y,ai]{I.~Tosta~e~Melo}
\collabauthor[j]{B.~Trocm{\'e}\,\orcidlink{0000-0001-9500-2487}}
\collabauthor[ao]{V.~Tsourapis\,\orcidlink{0009-0000-5616-5662}}
\collabauthor[aj]{C.~Tully\,\orcidlink{0000-0001-6771-2174}}
\collabauthor[ao]{E.~Tzamariudaki}
\collabauthor[au]{A.~Ukleja\,\orcidlink{0000-0003-0480-4850}}
\collabauthor[m]{A.~Vacheret}
\collabauthor[x]{V.~Valsecchi}
\collabauthor[ay,j]{V.~Van~Elewyck}
\collabauthor[k,l]{G.~Vannoye}
\collabauthor[b]{E.~Vannuccini}
\collabauthor[be]{G.~Vasileiadis}
\collabauthor[v]{F.~Vazquez~de~Sola}
\collabauthor[f,ae]{A. Veutro}
\collabauthor[x]{S.~Viola\,\orcidlink{0000-0001-9511-8279}}
\collabauthor[r,o]{D.~Vivolo}
\collabauthor[d]{A. van Vliet\,\orcidlink{0000-0003-2827-3361}}
\collabauthor[at]{L.~Voorend}
\collabauthor[ad,v]{E.~de~Wolf\,\orcidlink{0000-0002-8272-8681}}
\collabauthor[j]{I.~Lhenry-Yvon}
\collabauthor[l]{S.~Zavatarelli}
\collabauthor[x]{D.~Zito}
\collabauthor[e]{J.\,D.~Zornoza\,\orcidlink{0000-0002-1834-0690}}
\collabauthor[e]{J.~Z{\'u}{\~n}iga\,\orcidlink{0000-0002-1041-6451}}

\par\bigskip

% ---- Paste ONLY the \address[...] lines here ----
% (From your file: between "% ----- Start address list" and "% ----- End address list")

\address[a]{Universit{\`a} di Firenze, Dipartimento di Fisica e Astronomia, via Sansone 1, Sesto Fiorentino, 50019 Italy}
\address[b]{INFN, Sezione di Firenze, via Sansone 1, Sesto Fiorentino, 50019 Italy}
\address[c]{Universit{\'e}~de~Strasbourg,~CNRS,~IPHC~UMR~7178,~F-67000~Strasbourg,~France}
\address[d]{Khalifa University of Science and Technology, Department of Physics, PO Box 127788, Abu Dhabi,   United Arab Emirates}
\address[e]{IFIC - Instituto de F{\'\i}sica Corpuscular (CSIC - Universitat de Val{\`e}ncia), Catedr{\'a}tico Jos{\'e} Beltr{\'a}n, 2, 46980 Paterna, Valencia, Spain}
\address[f]{INFN, Sezione di Roma, Piazzale Aldo Moro, 2 - c/o Dipartimento di Fisica, Edificio, G.Marconi, Roma, 00185 Italy}
\address[g]{Universitat Polit{\`e}cnica de Catalunya, Laboratori d'Aplicacions Bioac{\'u}stiques, Centre Tecnol{\`o}gic de Vilanova i la Geltr{\'u}, Avda. Rambla Exposici{\'o}, s/n, Vilanova i la Geltr{\'u}, 08800 Spain}
\address[h]{Subatech, IMT Atlantique, IN2P3-CNRS, Nantes Universit{\'e}, 4 rue Alfred Kastler - La Chantrerie, Nantes, BP 20722 44307 France}
\address[i]{Universitat Polit{\`e}cnica de Val{\`e}ncia, Instituto de Investigaci{\'o}n para la Gesti{\'o}n Integrada de las Zonas Costeras, C/ Paranimf, 1, Gandia, 46730 Spain}
\address[j]{Universit{\'e} Paris Cit{\'e}, CNRS, Astroparticule et Cosmologie, F-75013 Paris, France}
\address[k]{Universit{\`a} di Genova, Via Dodecaneso 33, Genova, 16146 Italy}
\address[l]{INFN, Sezione di Genova, Via Dodecaneso 33, Genova, 16146 Italy}
\address[m]{LPC CAEN, Normandie Univ, ENSICAEN, UNICAEN, CNRS/IN2P3, 6 boulevard Mar{\'e}chal Juin, Caen, 14050 France}
\address[n]{North-West University, Centre for Space Research, Private Bag X6001, Potchefstroom, 2520 South Africa}
\address[o]{INFN, Sezione di Napoli, Complesso Universitario di Monte S. Angelo, Via Cintia ed. G, Napoli, 80126 Italy}
\address[p]{INFN, Sezione di Bologna, v.le C. Berti-Pichat, 6/2, Bologna, 40127 Italy}
\address[q]{Universit{\`a} di Bologna, Dipartimento di Fisica e Astronomia, v.le C. Berti-Pichat, 6/2, Bologna, 40127 Italy}
\address[r]{Universit{\`a} degli Studi della Campania "Luigi Vanvitelli", Dipartimento di Matematica e Fisica, viale Lincoln 5, Caserta, 81100 Italy}
\address[s]{E.\,A.~Milne Centre for Astrophysics, University~of~Hull, Hull, HU6 7RX, United Kingdom}
\address[t]{Czech Technical University in Prague, Institute of Experimental and Applied Physics, Husova 240/5, Prague, 110 00 Czech Republic}
\address[u]{Comenius University in Bratislava, Department of Nuclear Physics and Biophysics, Mlynska dolina F1, Bratislava, 842 48 Slovak Republic}
\address[v]{Nikhef, National Institute for Subatomic Physics, PO Box 41882, Amsterdam, 1009 DB Netherlands}
\address[w]{Aix~Marseille~Univ,~CNRS/IN2P3,~CPPM,~Marseille,~France}
\address[x]{INFN, Laboratori Nazionali del Sud, (LNS) Via S. Sofia 62, Catania, 95123 Italy}
\address[y]{INFN, Sezione di Catania, (INFN-CT) Via Santa Sofia 64, Catania, 95123 Italy}
\address[z]{University Mohammed V in Rabat, Faculty of Sciences, 4 av.~Ibn Battouta, B.P.~1014, R.P.~10000 Rabat, Morocco}
\address[aa]{Universit{\`a} di Salerno e INFN Gruppo Collegato di Salerno, Dipartimento di Fisica, Via Giovanni Paolo II 132, Fisciano, 84084 Italy}
\address[ab]{Universit{\`a} di Napoli ``Federico II'', Dip. Scienze Fisiche ``E. Pancini'', Complesso Universitario di Monte S. Angelo, Via Cintia ed. G, Napoli, 80126 Italy}
\address[ac]{Institute of Space Science - INFLPR Subsidiary, 409 Atomistilor Street, Magurele, Ilfov, 077125 Romania}
\address[ad]{University of Amsterdam, Institute of Physics/IHEF, PO Box 94216, Amsterdam, 1090 GE Netherlands}
\address[ae]{Universit{\`a} La Sapienza, Dipartimento di Fisica, Piazzale Aldo Moro 2, Roma, 00185 Italy}
\address[af]{Universit{\`a} di Bologna, Dipartimento di Ingegneria dell'Energia Elettrica e dell'Informazione "Guglielmo Marconi", Via dell'Universit{\`a} 50, Cesena, 47521 Italia}
\address[ag]{Cadi Ayyad University, Physics Department, Faculty of Science Semlalia, Av. My Abdellah, P.O.B. 2390, Marrakech, 40000 Morocco}
\address[ah]{University of the Witwatersrand, School of Physics, Private Bag 3, Johannesburg, Wits 2050 South Africa}
\address[ai]{Universit{\`a} di Catania, Dipartimento di Fisica e Astronomia "Ettore Majorana", (INFN-CT) Via Santa Sofia 64, Catania, 95123 Italy}
\address[aj]{Princeton University, Department of Physics, Jadwin Hall, Princeton, New Jersey, 08544 USA}
\address[ak]{INFN, Sezione di Bari, via Orabona, 4, Bari, 70125 Italy}
\address[al]{UCLouvain, Centre for Cosmology, Particle Physics and Phenomenology, Chemin du Cyclotron, 2, Louvain-la-Neuve, 1348 Belgium}
\address[am]{University of Granada, Department of Computer Engineering, Automation and Robotics / CITIC, 18071 Granada, Spain}
\address[an]{Friedrich-Alexander-Universit{\"a}t Erlangen-N{\"u}rnberg (FAU), Erlangen Centre for Astroparticle Physics, Nikolaus-Fiebiger-Stra{\ss}e 2, 91058 Erlangen, Germany}
\address[ao]{NCSR Demokritos, Institute of Nuclear and Particle Physics, Ag. Paraskevi Attikis, Athens, 15310 Greece}
\address[ap]{University Mohammed I, Faculty of Sciences, BV Mohammed VI, B.P.~717, R.P.~60000 Oujda, Morocco}
\address[aq]{Universit{\`a} degli Studi della Campania "Luigi Vanvitelli", CAPACITY, Laboratorio CIRCE - Dip. Di Matematica e Fisica - Viale Carlo III di Borbone 153, San Nicola La Strada, 81020 Italy}
\address[ar]{Western Sydney University, School of Science, Locked Bag 1797, Penrith, NSW 2751 Australia}
\address[as]{NIOZ (Royal Netherlands Institute for Sea Research), PO Box 59, Den Burg, Texel, 1790 AB, the Netherlands}
\address[at]{Leiden University, Leiden Institute of Physics, PO Box 9504, Leiden, 2300 RA Netherlands}
\address[au]{AGH University of Krakow, Al.~Mickiewicza 30, 30-059 Krakow, Poland}
\address[av]{LPC, Campus des C{\'e}zeaux 24, avenue des Landais BP 80026, Aubi{\`e}re Cedex, 63171 France}
\address[aw]{Tbilisi State University, Department of Physics, 3, Chavchavadze Ave., Tbilisi, 0179 Georgia}
\address[ax]{The University of Georgia, Institute of Physics, Kostava str. 77, Tbilisi, 0171 Georgia}
\address[ay]{Institut Universitaire de France, 1 rue Descartes, Paris, 75005 France}
\address[az]{Max-Planck-Institut~f{\"u}r~Radioastronomie,~Auf~dem H{\"u}gel~69,~53121~Bonn,~Germany}
\address[ba]{University of Sharjah, Sharjah Academy for Astronomy, Space Sciences, and Technology, University Campus - POB 27272, Sharjah, - United Arab Emirates}
\address[bb]{University of Granada, Dpto.~de F\'\i{}sica Te\'orica y del Cosmos \& C.A.F.P.E., 18071 Granada, Spain}
\address[bc]{School of Applied and Engineering Physics, Mohammed VI Polytechnic University, Ben Guerir, 43150, Morocco}
\address[bd]{University of Johannesburg, Department Physics, PO Box 524, Auckland Park, 2006 South Africa}
\address[be]{Laboratoire Univers et Particules de Montpellier, Place Eug{\`e}ne Bataillon - CC 72, Montpellier C{\'e}dex 05, 34095 France}
\address[bf]{Universit{\'e} de Haute Alsace, rue des Fr{\`e}res Lumi{\`e}re, 68093 Mulhouse Cedex, France}
\address[bg]{Universit{\'e} Badji Mokhtar, D{\'e}partement de Physique, Facult{\'e} des Sciences, Laboratoire de Physique des Rayonnements, B. P. 12, Annaba, 23000 Algeria}
\address[bh]{CSIC - Consejo Superior de Investigaciones Cientificas, ICM-CSIC - Instituto de Ciencias del Mar, Paseo Maritimo de la Barceloneta, 37-49, Barcelona, 8003 Spain}
\address[bi]{AstroCeNT, Nicolaus Copernicus Astronomical Center, Polish Academy of Sciences, Rektorska 4, Warsaw, 00-614 Poland}
\address[bj]{Charles University, Faculty of Mathematics and Physics, Ovocn{\'y} trh 5, Prague, 116 36 Czech Republic}
\address[bk]{Harvard University, Black Hole Initiative, 20 Garden Street, Cambridge, MA 02138 USA}
\address[bl]{School~of~Physics~and~Astronomy, Sun Yat-sen University, Zhuhai, China
}

\par\bigskip
\noindent\textsuperscript{*}Corresponding authors.
\endgroup